\documentclass[10pt,journal]{IEEEtran}

\usepackage{main_header}

\begin{document}

\title{Never Use Labels: Signal Strength-Based Bayesian Device-Free Localization in Changing Environments}

\author{Peter~Hillyard and
        Neal~Patwari
\IEEEcompsocitemizethanks{
\IEEEcompsocthanksitem P. Hillyard is with Xandem Technology, Salt Lake City, USA.  Contact e-mail: peter@xandem.com. N. Patwari is with the Department of Electrical and Computer Engineering, University of Utah, and with Xandem Technology, Salt Lake City, USA.}
}
\IEEEtitleabstractindextext{%
\begin{abstract}
Device-free localization (DFL) methods use measured changes in the received signal strength (RSS) between many pairs of RF nodes to provide location estimates of a person inside the wireless network. Fundamental challenges for RSS DFL methods include having a model of RSS measurements as a function of a person's location, and maintaining an accurate model as the environment changes over time. Current methods rely on either labeled empty-area calibration or labeled fingerprints with a person at each location. Both need to be frequently recalibrated or retrained to stay current with changing environments. Other DFL methods only localize people in motion. In this paper, we address these challenges by, first, introducing a new mixture model for link RSS as a function of a person's location, and second, providing the framework to update model parameters without ever being provided labeled data from either empty-area or known-location classes. We develop two new Bayesian localization methods based on our mixture model and experimentally validate our system at three test sites with seven days of measurements. We demonstrate that our methods localize a person with non-degrading performance in changing environments, and, in addition, reduce localization error by $\mathbf{11-51\%}$ compared to other DFL methods.
\end{abstract}
}

\maketitle
\IEEEdisplaynontitleabstractindextext
\IEEEpeerreviewmaketitle

\ifCLASSOPTIONcompsoc
\IEEEraisesectionheading{\section{Introduction}\label{S:introduction}}
\else
\section{Introduction}
\label{S:introduction}
\fi
\IEEEPARstart{W}{ireless} sensor networks have opened up many opportunities for detecting breaches in physical property, for home automation, and for remotely monitoring the health and activity of home-bound patients. These systems depend on knowing the location and presence of people in an area of interest. Previous research has shown how a person, without an RF tag, can be localized through walls by processing received signal strength (RSS) measurements between many pairs of statically deployed RF nodes. This tag-less based localization technology is known as device-free localization (DFL) \cite{youssef2007challenges}.

DFL in indoor environments presents many significant challenges. First, the links' RSS are non-stationary in changing environments. Consequently, DFL methods that require an empty-area calibration \cite{wilson2010rti,kaltiokallio2012channels,savazzi2016assisted,savazzi2014bayesian} or fingerprint training \cite{xu2012classification} will need frequent recalibration or retraining to adjust to a changing environment. In contrast, online calibration methods \cite{wilson2011vrti,zhoa2013krti} quickly adjust to a changing environment; yet, they can only locate people while they are in motion. For real-world installations of DFL, both DFL types may be unacceptable. For example, a home-bound patient may find it very inconvenient to leave their home to provide a DFL system an empty-area calibration period. They may find it equally inconvenient to retrain their system by labelling training data with true locations. Also, smart-home systems must estimate occupancy, not only motion, in order to control lighting, heating and cooling systems.

One contribution of this paper is that we develop and validate a localization system that addresses the drawbacks of traditional empty room calibration, fingerprint training, and online calibration methods. The significance of our contribution is illustrated by stepping through the installation and unlabelled training process. In our system, which is built for a single occupant building, nodes are first installed around the building. After nodes have been deployed, the occupant walks around in the building during a system setup period. Alternatively, in the case of the home-bound patient, the node installer walks around the building while the home-bound patient remains stationary. During system setup, unlabelled training data is fed to the system and used to estimate model parameters. After system setup is complete, the system localizes the building occupant. Unlabelled training data continues to be fed to the system after system setup to adapt to non-stationary RSS measurements in changing environments, a method we call \textit{continuous recalibration}. Using unlabelled training data from a person naturally moving inside the building provides a more convenient way of training the system and to locate both a stationary and moving person. 

Other continuous recalibration methods with unlabelled training were presented in \cite{kaltiokallio2017arti,bocca2013assisted}. However, these methods assumed that an empty-area calibration was performed before runtime, unlike our methods in which no empty-area calibration is assumed. In \cite[\S VI-D]{kaltiokallio2017arti}, an approach is proposed in which, instead of performing an empty-area calibration, it is estimated as a person walks inside the building, however, the localization performance when using this idea is not reported. This paper quantifies localization performance of a method in which a person moving inside the building provides unlabelled training data. Our experiments include intentional and unintentional changes to the background that allow us to evaluate our algorithms over time as the environment changes. Further, we note that this method is complementary to the developments of \cite{kaltiokallio2017arti}, which performs filtering and smoothing to both image and coordinate estimates to refine the person's track over a period of time so that model parameters can be as accurate as possible. Our method primarily estimates areas where a person is very likely \emph{not} located as a means to improve knowledge of link model parameters.

A second challenge with indoor DFL is that, because of multipath fading, it is difficult to model the effect of a person's location on the measured RSS of a link. With a fine resolution of fingerprint locations, fingerprint training can capture this relationship in a particular environment by measuring the relative frequency of RSS measurements at each location \cite{seifeldin2013nuzzer}. While fingerprinting can be very accurate in localizing people, the fingerprints must be frequently retrained to stay current in changing environments \cite{mager2015fingerprint}. Alternative DFL methods like radio tomographic imaging (RTI) \cite{wilson2010rti}, Bayesian methods \cite{hillyard2016boundary}, and particle filters \cite{chen2011montecarlo,wilson2012skew,zheng2012foreground,savazzi2014bayesian} provide more flexibility for DFL because the relationship between measured RSS and a person's location is modeled \emph{a priori}.

Fundamental to model-based DFL is the idea that a link is \emph{affected}, i.e., has significant measured changes in RSS, when a person is on the link line. The link line is the imaginary line segment connecting the link's nodes. In contrast, when the link is \emph{unaffected}, i.e., has very little measured change in RSS, the person tends to be off of the link line. Model-based DFL methods have been built around the idea that the link is affected only when a person is inside an ellipse whose foci are the nodes of the link \cite{wilson2010rti,kaltiokallio2012channels,savazzi2014bayesian}. In reality, the link can be affected even when a person is far from the link line, or unaffected when a person is on the link line. These model inaccuracies confound Bayesian DFL methods.

As a second contribution of this paper, we develop a new mixture model where a link may be affected or unaffected no matter the person's position, but with probabilities that are a function of the person's distance from the link line. In our system, we learn RSS distribution parameters for both the affected and unaffected state of each link and we do so with unlabelled measurements. The weights in our mixture model are derived from a spatial model such that the affected RSS distribution is weighted more when a person is on the link line and weighted less the further the person is from the link line. 

We incorporate our new mixture model in two Bayesian localization methods we develop which we refer to as maximum likelihood localization (MLL) and hidden Markov model localization (HMML). These two methods differ from other Bayesian localization methods \cite{wilson2012skew,savazzi2016assisted,savazzi2014bayesian} by incorporating randomness in the affected and unaffected state. MLL and HMML both compute the probability of observing the measured RSS given a person's location. Adding a temporal property to localization, HMML extends MLL by estimating the current location based on the previous location. In that both MLL and HMML operate on the same mixture and spatial models, and only differ in their temporal properties, we refer to them generally as model-based probabilistic localization (MPL). However, we differentiate between the localization method used in MPL as either MLL or HMML.

We experimentally validate HMML and MLL at three separate sites and with over 7 days of measured RSS data. We demonstrate that MPL does not need an empty room calibration or fingerprint training period, that it adapts to changes in RSS due to a changing background, and that it is capable of localizing a stationary person. We compare HMML and MLL to an RTI method which uses empty room calibration \cite{kaltiokallio2012channels}, to an RTI method that uses online calibration \cite{zhoa2013krti}, and to a Bayesian linear discriminant analysis method \cite{xu2012classification} which localizes with the help of a database of labelled fingerprint measurements. We show that HMML and MLL can match or decrease the localization error by $11-55\%$ compared to these other DFL methods. Additionally, by reducing missed detection errors by orders of magnitude and reducing the false alarm rate by a factor of two to four, we show that we can track stationary targets despite changing environments.

\section{Related Work}

The ability to locate a person indoors using sensors has changed the way we think about security, home automation and smart homes, and aging in place. Some of these sensing systems include: cameras that detect changes in pixel values caused by a person's presence \cite{anderson2006falls}; pyroelectric sensors that detect and locate changes in thermal radiation due to a person's presence \cite{hao2006pyroelectric}; and 
vibration sensors to localize vibrations from a person walking \cite{mirshekari2016characterizing}. 
Cameras and infrared sensors cannot sense through material opaque to visible light, and vibration sensors must be sensitive enough to detect vibrations on the inside of the home while ignoring ambient vibrations. Our MPL solution, like other RF solutions, is a more appropriate choice for whole-home, through-wall sensing since RF can sense through walls, smoke, and in any lighting condition.

RF sensing systems perform localization in a variety of ways. Ultra-wideband radios can be used in multi-static radar to measure the time-of-flight between pulse transmissions and received reflections caused by moving people \cite{mccracken2014bistatic}. The time-of-flight is proportional to the distance between the reflector and the transmitter and receiver which is used to localize a person. Ultra-wideband radios have also been used measure changes in the line-of-sight power and then perform tomography with those measurements \cite{beck2016uwb}. A person's presence has also been shown to create significant changes in the amplitude of subcarriers in PHY layer measurements of commodity WiFi cards. These changes have been used in a fingerprint classification method to localize a person \cite{nasser2013monophy}. Instead of relying on fingerprinting, systems that use WiFi channel state information could alternatively apply the methods developed in this paper to perform localization.

Our MPL method is complementary to methods that process RSS to perform DFL including particle filters \cite{wilson2012skew,chen2011montecarlo}, fingerprint classification \cite{xu2012classification,mager2015fingerprint}, and RTI \cite{wilson2010rti,kaltiokallio2012channels}. These methods, however, either need a person to stand at several locations, to have the area completely vacant for a short period \cite{wilson2010rti,kaltiokallio2012channels,saeed2014ichnaea,bocca2013assisted}, or to have the person continuously moving in order to perform localization \cite{edelstein2013calibration,zhoa2013krti}. What sets our work apart is that MPL can perform DFL without fingerprinting, without a vacant area, and can localize stationary people.

DFL methods vary in how RSS measurements are used to estimate location. In fingerprint-based localization, a person stands at many locations in the area of interest while the statistics of the RSS distributions are recorded \cite{mager2015fingerprint,xu2012classification}. During testing, the likelihoods or Bayesian probability of measuring the observed RSS is computed for each fingerprint from which the estimated location is derived. With enough fingerprint locations, fingerprint-based localization captures the hard-to-predict RSS distribution of a link as a function of a person's location. However, this comes at an unsustainable cost of frequently retraining fingerprints to stay current with changing environments. MPL seeks to provide a highly accurate localization system, like that of fingerprinting, but by doing so with a model that provides more flexibility in changing environments.

In RTI methods, a spatial model is assumed which indicates where a person's presence will cause a change in RSS  \cite{wilson2010rti,kaltiokallio2012channels}. An image of the most likely locations a person was is formed based on the change in RSS observed on the links in the network. A more recent version of RTI has been developed to learn the parameters of the spatial model for each link with unlabelled data \cite{kaltiokallio2017arti}. Additionally, the empty room calibration measurements are continuously recalibrated as a person moves inside the area of interest to stay current in changing environments. Furthermore, the change in the RSS on a link is weighted as a function of the excess path length of the person's location and the link. MPL provides an alternative approach to localization by computing the probability of observing RSS measurements based on a person's location. Furthermore, our system performs continuous recalibration not only when a person is moving inside the area of interest but also when the area of interest is vacant. Continuous recalibration in both cases is necessary to avoid nuisance alarms when the area of interest is vacant.

Another variation of DFL is particle filtering. As in MPL, particle filters assume RSS distributions, one for when a person is on a link line, and one for when they are off of the link line \cite{wilson2012skew,zheng2012foreground,chen2011montecarlo,savazzi2014bayesian,savazzi2016assisted}. Particles are then drawn from a Gaussian distributions and are said to be drawn from the affected RSS distribution when their excess path length to a link are less than some threshold. Otherwise they are drawn from the unaffected RSS distribution. MPL differs from this approach in that we do not place 0 or 1 weights to the affected and unaffected RSS distribution, but soft weights that are a function of the person's excess path length to a link. This approach inserts some uncertainty in the model to account for the reality that a link may not be affected even when a person is standing on the link line or that a link is affected when the person is far from the link line. As another point of differentiation, our methods, unlike \cite{savazzi2014bayesian,savazzi2016assisted} actually implement and validate a method that tracks both a stationary target and a target in motion without using labelled calibration data.

\section{Methods}
\begin{figure}
\centering
\includegraphics[width=0.46\textwidth]{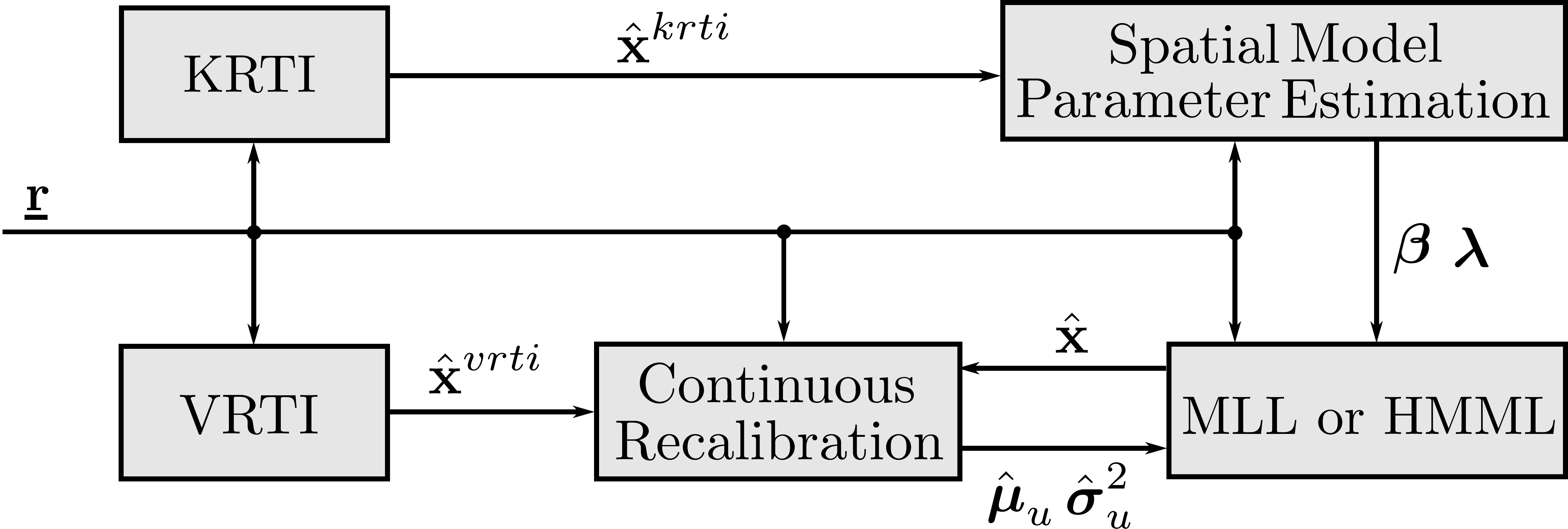}
\caption{Block diagram of model-based probabilistic localization (MPL)}
\label{F:block_diagram}
\vspace{-1em}
\end{figure}

In this section, we describe the fundamental components of MPL generally and of MLL and HMML specifically. The components of MPL are shown in the block diagram in Fig.~\ref{F:block_diagram} and include: a one-time estimation of the weights of the mixture models that relate RSS to an occupied location; a lightweight online RTI method that runs in tandem with either MLL or HMML to provide a location of a moving person; a continuous recalibration block that continuously re-estimates the parameters of the links' affected and unaffected distributions; and a block where MLL or HMML is implemented. MLL and HMML compute the probabilities of observing the RSS measurements on the links given the person's location. We describe each of these components in more detail in the following sections.

%
%
%
%
%
%
%
%
%
%
%
%
%
\subsection{Equipment and Measurements}
In this paper, our wireless measurements are made using Texas Instruments CC2531 dongles that communicate using IEEE 802.15.4 in the 2.4 GHz ISM band. We deploy $N$ nodes around the area of interest. The nodes are programmed to take turns transmitting a packet on a 802.15.4 channel during dedicated time slots using TDMA and a token-ring passing protocol \cite{multispin}. This protocol is repeated on a predefined set of 802.15.4 channels.

As each node transmits on each channel, a separate node logs the RSS, also called the received power in decibel units, between each pairwise node. We denote the RSS measured on link $l=(i,j,c)$ formed by transmitting node $i$ and receiving node $j$ on channel $c$ as $r_l$. The RSS is typically a discrete-valued measurement, and we denote its possible values as $\mathcal{S}_r$. $\mathcal{S}_r$ also includes $\oslash$, the event that there was a missed packet and as such RSS was not measured. We observe a vector $\mathbf{r} = [r_1,r_2,\ldots,r_L]$ on $L$ links. 

%
%
%
%
%
%
%
%
%
%
%
%
%
\subsection{Mixture Model} \label{S:mixture_model}
As in many model-based DFL methods, MPL adopts the idea that a link is either in an \textit{affected} state or an \textit{unaffected} state \cite{hillyard2016boundary,wilson2012skew,zheng2012foreground}. However, the novelty in this model is that, given the person's location, the state of the link is not known \emph{a priori}. In contrast, some models state that a link deterministically is affected when a person is present in an ellipse whose foci are the node coordinates of the link and is unaffected when the person is outside of the ellipse \cite{wilson2012skew,wilson2010rti,wilson2011vrti,kaltiokallio2012channels}. Another model states that a person's presence in a voxel intersected by the link line causes the link to be deterministically affected \cite{kanso2009compressed}. However, by virtue of the random nature of the multipath radio channel, any deterministic model for the state of a link as a function of person location is bound to be inaccurate. Further, if pixels are large (to keep computation time low), there may be, within the pixel, positions in which the person affects a link as well as positions in which they do not affect it. We develop a mixture model that places some uncertainty on the whether a link is affected or unaffected by a person's presence in a pixel.

\begin{figure}
\centering
\includegraphics[width=0.7\columnwidth]{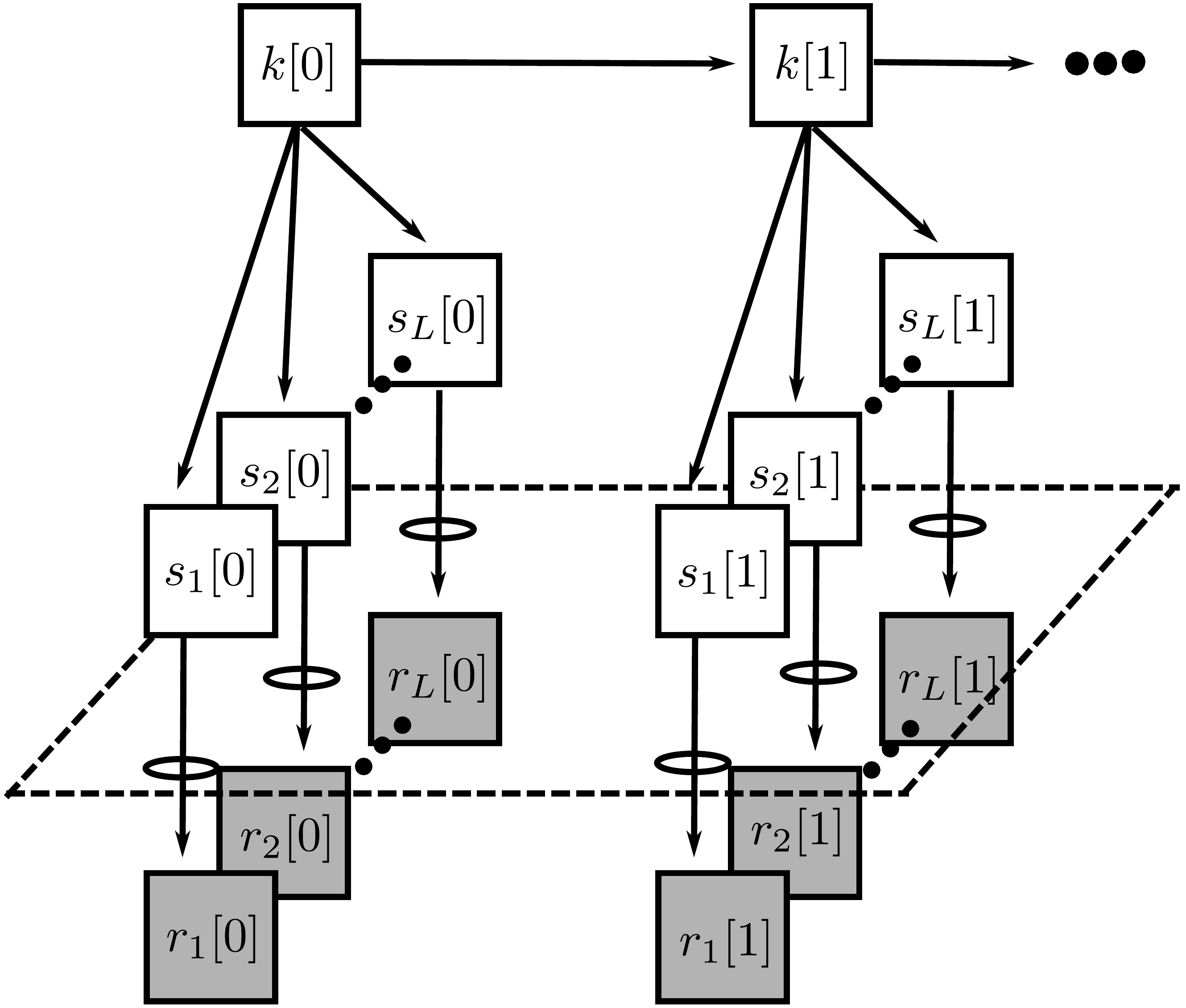}
\caption{Bayesian graphical model}
\label{F:mixture}
\vspace{-1em}
\end{figure}

Our mixture model follows the diagram shown in Fig.~\ref{F:mixture}
First, assume that a person is present at one of $P+1$ grid coordinates $\mathbf{x}^{grid}_k$ for pixel $k \in \{0,\ldots,P\}$. In one time step, a person can transition from one grid point to any other grid point that can be reached by a person moving at the maximum assumed velocity (we use 3 m/s unless otherwise stated). While a person is located in pixel $k$, a link $l$ has a state $s_l \in \{ a,u\}$ where $a$ is the affected state and $u$ is the unaffected state.  Our link state probability, given the person's location, is denoted $p(s_l \mid k)$. The probability a link is affected is an exponentially decaying function of excess path length 
\begin{equation} \label{E:conditional_probs}
p(s_l = a \mid k) = \beta_l \cdot e^{-\delta_{l,k}/\lambda_l}
\end{equation} where the excess path length of location $k$ with respect to link $l$ is 
$\delta_{l,k} = d(\mathbf{x}^{grid}_k,\mathbf{x}^{tx}_l) + d(\mathbf{x}^{grid}_k,\mathbf{x}^{rx}_l) - d(\mathbf{x}^{tx}_l,\mathbf{x}^{rx}_l)$, where $d(\mathbf{x},\mathbf{y})$ is the Euclidean norm between $\mathbf{x}$ and $\mathbf{y}$, $\mathbf{x}^{tx}_l$ is the coordinate of link $l$'s transmitter, $\mathbf{x}^{rx}_l$ is the coordinate of link $l$'s receiver, and $\beta_l$ and $\lambda_l$ are parameters we will estimate. The probability that a link is unaffected by a person standing in pixel  $k$ is $p(s_l = u \mid k) = 1 - p_l(s_l = a \mid k)$.

If a link is affected, the RSS on link $l$, $r_l$, is  generated from the conditional distribution $p(r_l \mid s_l = a)$. The probability of observing $r_l$ in the affected state is weighted by $p(s_l=a \mid k)$, the probability that the link is affected given the person is standing in pixel $k$. If a link is unaffected, $r_l$ is randomly generated from the conditional distribution $p(r_l \mid s_l = u)$. The probability of observing $r_l$ in the unaffected state is weighted by $p(s_l = u \mid k)$. Via Bayes' Law, we can see that $r_l$ given a person at position $k$ is generated from the mixture model
\begin{equation} \label{E:mixture_model}
p( r_l \mid k ) = \sum_{s' = a,u} p(s_l = s' \mid k) \cdot p(r_l \mid s_l = s').
\end{equation}

Assuming link RSS measurements are independent, the likelihood that the person is in pixel $k$ given $\mathbf{r}$ is
\begin{equation} \label{E:likelihoods}
p_k( \mathbf{r} ) = \prod_{l=1}^L p( r_l \mid k ).
\end{equation} However, this product may not be able to be represented by modern computers when $L$ is large. To avoid these issues, we compute log probabilities first and then convert them back into probabilities as $p_k( \mathbf{r} ) = \exp \left\{ \sum_{l} \log p( r_l \mid k ) - \psi \right\}$ where $\psi = \max_k \sum_{l} \log p( r_l \mid k )$. For MLL, the estimated location of the person is found from
\begin{equation}
\hat{\mathbf{x}}^{mll} = \argmax_{0 \leq k \leq P} p_k( \mathbf{r} ).
\end{equation} 

The HMML solves for the most likely location given a history of $\mathbf{r}$ observations by inductively computing a forward probability vector, $\alpha_k[t]$, at time $t$ for each grid coordinate $k=\{0,\ldots,P\}$. The value of $\alpha_k[t]$ is the joint probability of current state $k$ and all link RSS measurements $\mathbf{r}$ through time $t$ \cite{rabiner1989hmm}.
The HMML estimates the current location of the person as
\begin{equation}
\hat{\mathbf{x}}^{hmml}[t] = \argmax_{0 \leq k \leq P} \alpha_k[t].
\end{equation} The probability that the initial pixel of the person is $k$ is denoted $\pi_k$. The forward algorithm initializes $\alpha_k[1] = \pi_k p_k(\mathbf{r}[1])$ where $\mathbf{r}[1]$ is the first measured RSS vector, and then computes $\alpha_k[t+1] = \left[ \sum_{w=0}^{P} \alpha_w[t] p_{wk} \right] \cdot p_k(\mathbf{r}[t+1])$ for each $t>1$ and for $ 0 \leq k \leq P$ where $p_{wk}$ is the probability that a person transitions from pixel $w$ to pixel $k$ in one time step. For reference, the grid coordinates are evenly distributed. One coordinate represents the out-of-the-area coordinate which we denote $\mathbf{x}^{grid}_{P} = [\infty,\infty]$. For this out-of-the-area coordinate, we manually set $p_l(a \mid \mathbf{x}^{grid}_{P}) = \num{1e-3}$.

%
%
%
%
%
%
%
%
%
%
%
%
%
\subsection{Conditional RSS Distributions} \label{S:dist_models}
\begin{figure}
\centering
\includegraphics[width=0.4\textwidth]{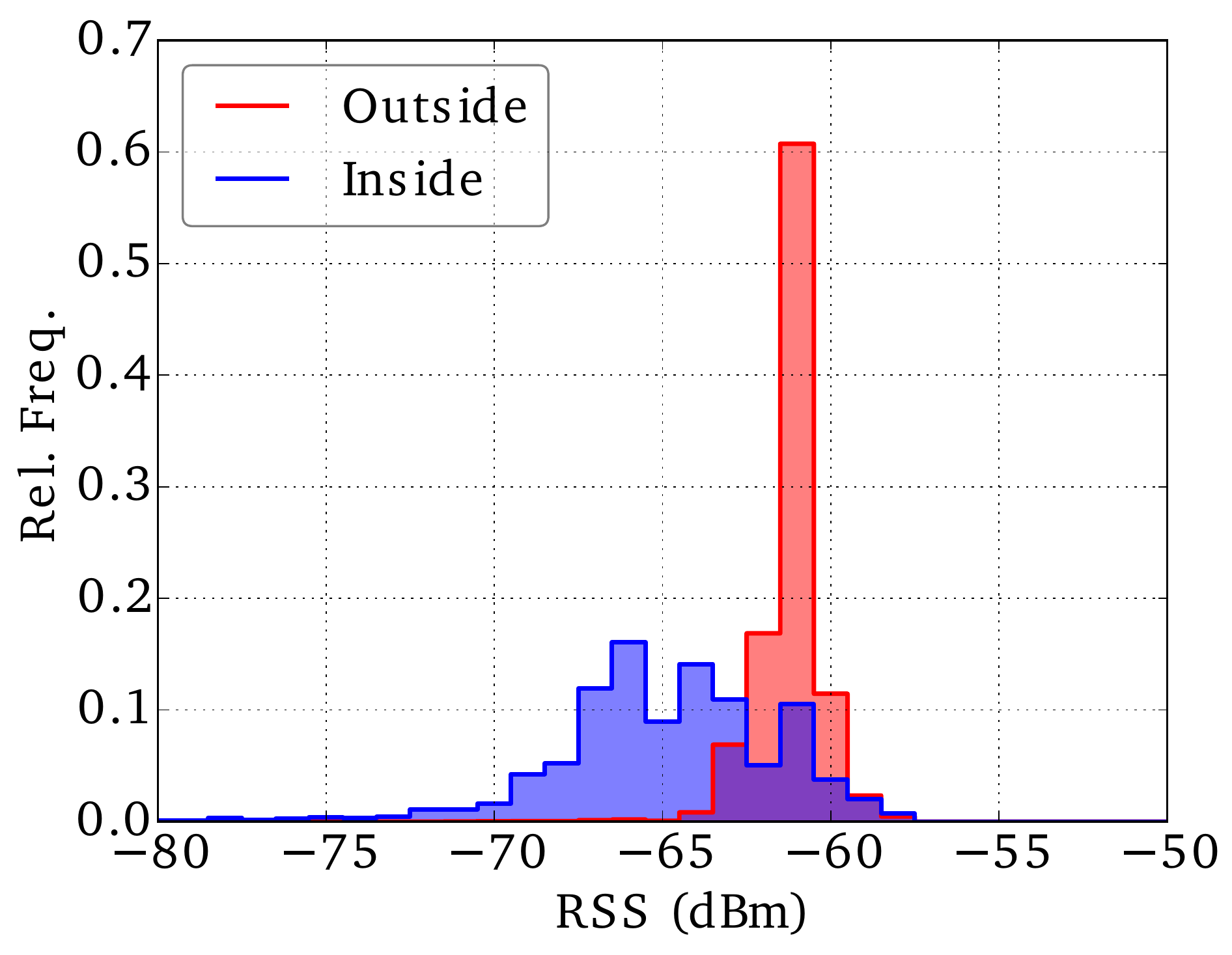}
\caption{Distribution of RSS on one link when a person is inside or outside an ellipse whose foci are the transmitter and receiver coordinates of a link. This link represents the behavior of the majority of links.}
\label{F:aff_unaff_dists}
\vspace{-1em}
\end{figure}

In this work, we have adopted the idea that a link is either in an affected or unaffected state. We perform our own experiments to support this claim. In this experiment, a person walks around a room at known times and at known coordinates while we record RSS measurements. We show in Fig.~\ref{F:aff_unaff_dists} one link's distribution of RSS when a person is far from the link line and when the person is on or near the link line. While the RSS does not assume these distributions in all cases, we use this link to represent the behavior of the majority of links.
 The affected and unaffected RSS conditional distributions can be modeled as skew-Laplace \cite{wilson2012skew}, or Ricean \cite{ghaddar2007cylinder}, but we sacrifice model accuracy by using a Gaussian model for simplicity. 
A normal distribution for RSS in decibel units has also been adopted in \cite{xu2012classification,zheng2012foreground}. The mean and variance of the unaffected distribution we denote as $\mu_{l,u}$ and $\sigma^2_{l,u}$ where the subscript $u$ specifies unaffected. The mean and variance of the affected distribution we denote as $\mu_{l,a}$ and $\sigma^2_{l,a}$ where the subscript $a$ specifies affected. 

For link $l$, we estimate the mean and variance of both distributions using RSS measurements when there is evidence that the link is unaffected. We describe in Section \ref{S:continuous_recal} how we decide when RSS is measured when the link is unaffected, but for now, we create a FIFO buffer of length $B$ for link $l$. When a link is unaffected, we add $r_l$ to the buffer. When a measurement is added to the buffer, we compute both the sample mean and the sample variance of the buffer which we save as $\hat{\mu}_{l,u}$ and $\hat{\sigma}^2_{l,u}$ respectively. When there are no changes to objects in the background environment, we anticipate $\hat{\mu}_{l,u}$ to be about the same as $\mu_{l,u}$. So, we only perform the update $\mu_{l,u} \leftarrow \hat{\mu}_{l,u}$ and $\sigma^2_{l,u} \leftarrow \hat{\sigma}^2_{l,u}$ when $|\mu_{l,u}-\tilde{\mu}_{l,u}| > 1$ dBm. 

From Fig.~\ref{F:aff_unaff_dists}, we also observe that the mean of the unaffected histograms is a few dBm greater than the affected histogram's mean. Also, the variance of the affected histogram is larger than the unaffected variance. In our model, we use these observations to also estimate the mean and variance of the affected distribution by $\mu_{l,a} \leftarrow \hat{\mu}_{l,u} - \Delta$ and $\sigma^2_{l,a} \leftarrow \eta\hat{\sigma}^2_{l,u}$. We have found that $\Delta = 3$ dBm and $\eta=2.5$ are appropriate parameters to use for indoor settings. We also note that to estimate $\hat{\sigma}^2_{l,u}$, we use the maximum of the sample variance of the buffer and a minimum constant $\omega^2>0$. Due to quantization of RSS, the sample variance may be zero even though the true real-valued received power would have had a positive variance. We impose a minimum variance of $\omega^2>0$ to avoid numerical instability. We have found that $\omega=0.75$ is an appropriate value for this application.

In reality, the effects of multipath fading would mean each link would have a unique $\Delta$ and $\eta$ instead of the fixed value we use. Although fixed in this paper, the values of $\Delta$ and $\eta$ for all links captures the general RSS response, which is a drop in RSS and an increase in variance, when a person is nearby the link line. The choice to estimate each link's $\Delta$ and $\eta$ from noisy target location estimates would add another layer of high-dimension estimation which we chose not to explore in this paper.

When the mean and variance of a link's unaffected and affected distributions have been re-estimated, we recompute their RSS mass functions as
\begin{equation} \label{E:log_normal_unaffected}
p(r_l \mid u)  = \left\{\begin{array}{ll} \epsilon, & r_l=\oslash \\ \max\left\{ \epsilon, \frac{1}{\gamma} \mathcal{N} (r_l; \mu_{l,u}, \sigma^2_{l,u}) \right\}, & r_l \neq \oslash \end{array}\right. 
\end{equation} 
and
\begin{equation} \label{E:log_normal_affected}
p(r_l \mid a)  = \left\{\begin{array}{ll} \epsilon, & r_l=\oslash \\ \max\left\{ \epsilon, \frac{1}{\gamma} \mathcal{N} (r_l; \mu_{l,a}, \sigma^2_{l,a}) \right\}, & r_l \neq \oslash \end{array}\right.
\end{equation}
where $\gamma$ is constant such that the pdf (probability distribution function) sums to one, and $\epsilon>0$ is a small-valued lower bound on the probability value away from zero. The use of the minimum probability $\epsilon$ is due to the fact that, in practice, we may observe values far from the mean more often than described by equations (\ref{E:log_normal_unaffected}) and (\ref{E:log_normal_affected}) because temporal fading does not always fit the log-normal distribution \cite{hashemi1994temporal}. Using a small value $\epsilon$ conveys the model uncertainty and avoids numerical issues with very low probabilities in the likelihood computations. We use $\num{1e-5}$ in this work. 

%
%
%
%
%
%
%
%
%
%
%
%
%
\subsection{Spatial Model} \label{S:spatial_model}
The approach with many DFL methods is to say a link is affected only when a person is standing inside an ellipse whose foci are the coordinates of the link's nodes \cite{wilson2010rti}. Instead of setting a strict elliptical boundary on when the link is affected and unaffected, other models have used a decaying elliptical model where changes in RSS on a link are weighted according to the person's excess path length to the link \cite{kaltiokallio2017arti}. 

\begin{figure}
\centering
\includegraphics[width=0.38\textwidth]{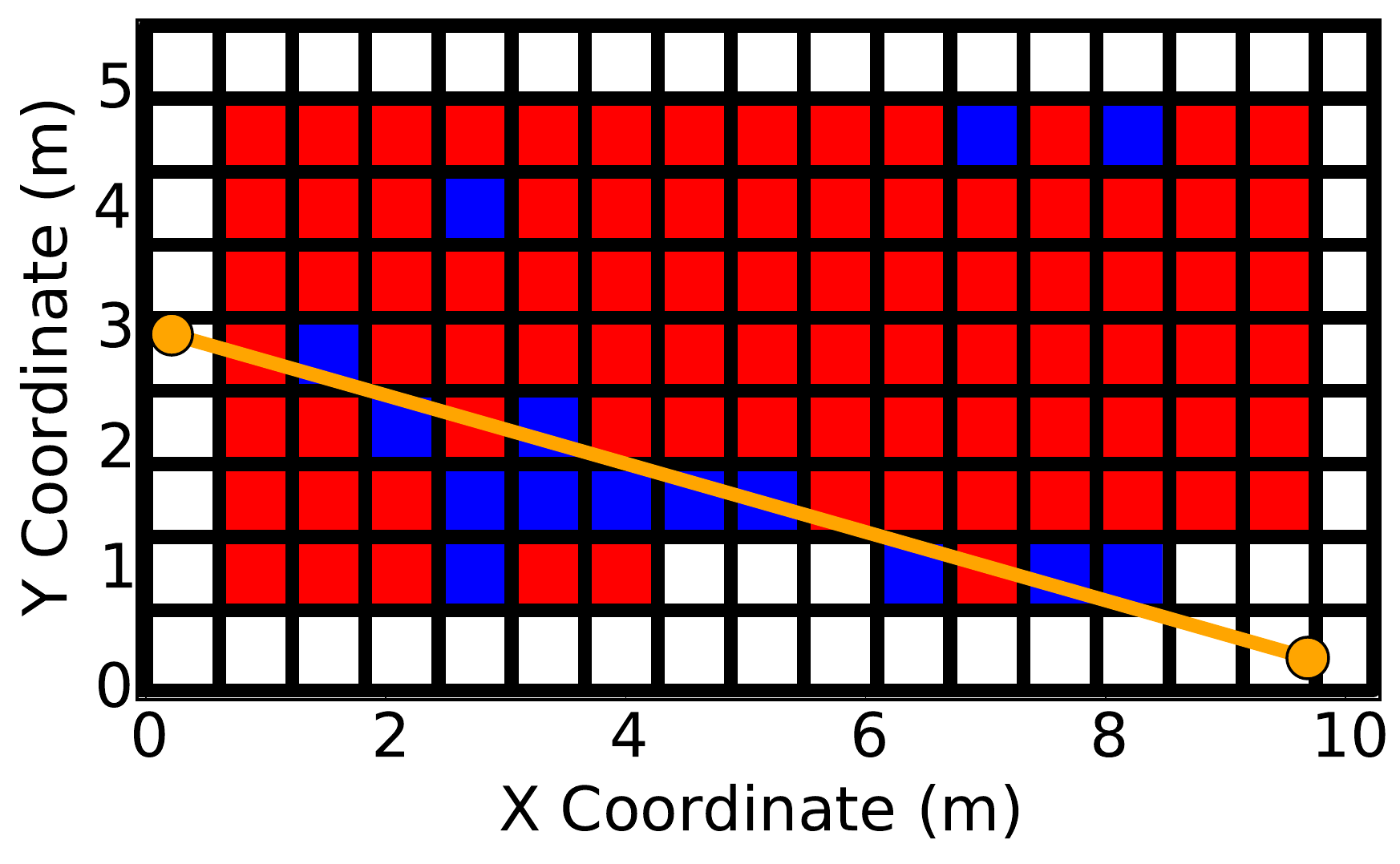}
\caption{The absolute difference between the mean RSS during empty room and the mean RSS when a person occupied each square. Red squares represent an absolute difference of 2 dBm or less. Blue squares represent an absolute difference greater than 2 dBm. The white squares were never occupied. The nodes and link line are shown in orange.}
\label{F:aff_unaff_pixels}
\vspace{-1em}
\end{figure}

In MPL, we adopt a similar decaying elliptical model that we base off of an experiment that we perform in an empty classroom. In this experiment, a person moves inside many $1.22 m^2$ areas for 30 seconds each, during which time RSS for many links are measured and recorded. An additional 30 seconds of RSS is recorded when the person is not in the area of interest. In post-processing, we find the mean RSS for each link and for each location the person occupied, including when the person stood outside the area of interest. In Fig.~\ref{F:aff_unaff_pixels}, we show the absolute difference in mean RSS when a person occupies each $1.22 m^2$ area and the mean RSS when the area of interest in vacant for link $l$.
 We threshold the image so that the absolute differences that are greater than 2 dBm are shown in blue, and smaller differences are shown in red. White squares are never occupied. 

What Fig.~\ref{F:aff_unaff_pixels} shows is which areas experience a change in mean RSS when a person occupies the area. In Fig.~\ref{F:aff_unaff_pixels}, we observe that, in general, areas near the link line tend to result in a decrease in RSS. Areas that are further away tend to experience small differences in RSS. However, we also observe that some locations show no measured change in RSS even when the person is on the link line. Additionally, when a person is very far from the link line, the link's RSS can significantly change in mean. A simple elliptical model does not capture the uncertainties due to multipath fading. The spatial model in equation (\ref{E:conditional_probs}) creates some uncertainty in our mixture model so that there is always a small but significant probability that an RSS measurement was drawn from either the affected or unaffected conditional distribution. When the person is near the link line, the affected distribution is weighted more heavily than the unaffected distribution. The choice of $\beta_l$ and $\lambda_l$ gives us some control over how the weights in the model are selected so that we can adjust to the different fading characteristics of each link. 

%
%
%
%
%
%
%
%
%
%
%
%
%
\subsection{Estimating Spatial Model Parameters} \label{S:spatial_param_estimation}
In this section we describe how we estimate the mixture model parameters $\beta_l$ and $\lambda_l$ for each link $l$. To accomplish this, our goal is to estimate $\beta_l$ and $\lambda_l$ such that our mixture model closely matches the distribution of RSS measurements as a function of the excess path length of the person's location and link $l$. This estimation process refers to the KRTI and parameter estimation block seen in Fig.~\ref{F:block_diagram}. KRTI is an online DFL method that does not require an empty room calibration period \cite{zhoa2013krti}. We choose KRTI because of its relatively low computational complexity and its highly accurate localization capability. KRTI updates a long and short term RSS histogram with every new RSS measurement. The difference between these two histograms is computed using the kernel distance. The differences from all of the links are then used to form an image and estimate the location of a person. During a training period a person walks inside the area of interest. During this time, KRTI provides an estimated location, $\hat{\mathbf{x}}^{krti}$, for each $r_l$. We store all $<r_l,\delta^{krti}_{l}>$ tuples where $\delta^{krti}_{l}$ is the excess path length between $\hat{\mathbf{x}}^{krti}$ and link $l$. 

After the training period is complete, we first estimate the mean and variance of the unaffected distribution using, respectively, the median and median absolute deviation (MAD) of the RSS during the training period. We use these statistics to estimate the mean and variance of the unaffected RSS distribution. The median and MAD ignore RSS measurements that fall far from the true unaffected mean and robustly estimate the mean and variance. Additionally, we multiply the MAD by $1.48$ and square the value to make it an estimate of the variance for Gaussian data \cite{rousseeuw1993mad}. Once the unaffected mean and variance have been estimated, we apply the same shift to the mean and scale to the variance to get the affected mean and variance as described in Section \ref{S:dist_models}. 

Instead of basing the affected distribution parameters on the unaffected distribution parameters, an algorithm could discriminate between times when a link is in either the affected or unaffected state and then directly estimate the parameters for the affected and unaffected state. However, we choose to not pursue this approach because it would require a long training period to ensure that there were a sufficient number of RSS measurements from each state to estimate the distribution parameters. Additionally, a person may not be physically capable of reaching locations where they would be affecting a link, and therefore, there would be no RSS measurements to estimate the affected distribution parameters.

\begin{figure}
\centering
\includegraphics[width=0.475\textwidth]{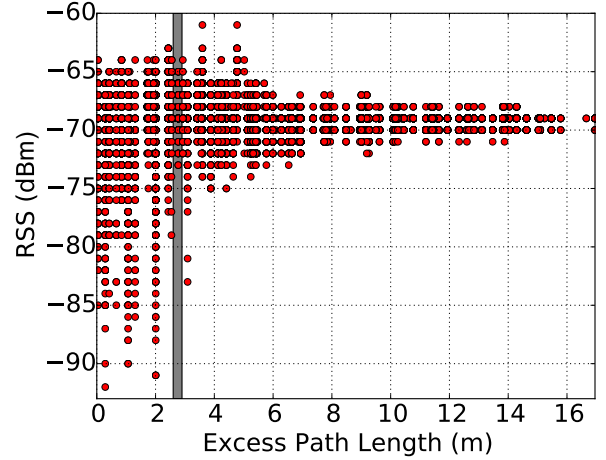}
\caption{Measured RSS as a function of excess path length, computed using the estimated location from the KRTI block seen in Fig.~\ref{F:block_diagram}. One group of RSS measurements with the same excess path length is shown in the gray box.}
\label{F:rss_vs_epl}
\end{figure}

\begin{figure}
\centering
\includegraphics[width=0.36\textwidth]{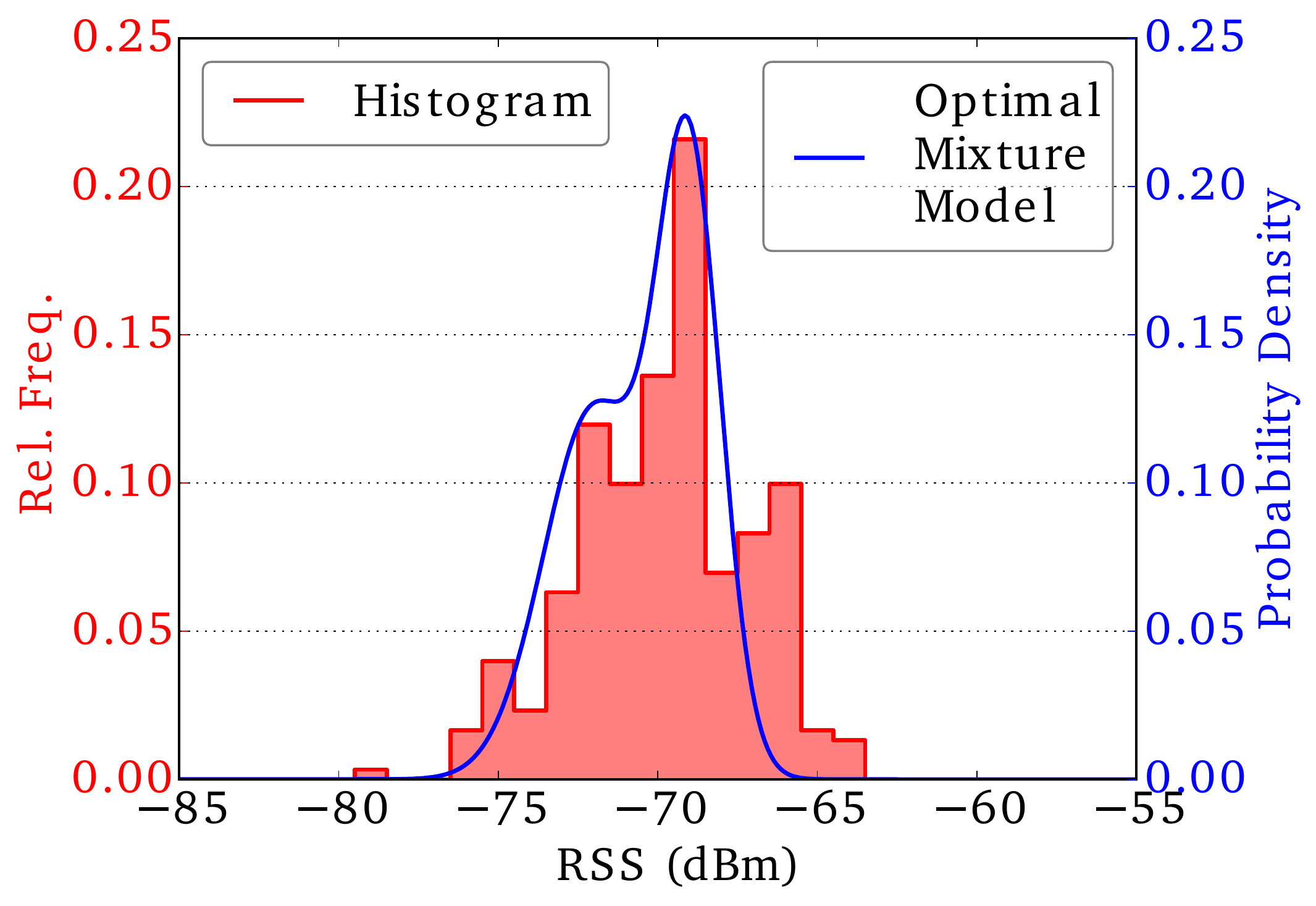}
\caption{Histogram of RSS within the gray box of Fig.~\ref{F:rss_vs_epl}. The mixture model is found using equation \ref{E:b_minimize} for this histogram and is overlaid.}
\label{F:rss_dist_opt_mix}
\vspace{-1em}
\end{figure}

After the training period is complete and the affected and unaffected RSS distribution parameters are estimated, we turn to the RSS, excess path length tuples previously mentioned. An example of the tuples for one of the links is shown in Fig.~\ref{F:rss_vs_epl}.
We next divide $r_l$ into bins according to excess path length $\delta^{krti}_{l}$. We choose to bin all tuples $<r_l,\delta^{krti}_{l}>$ into groups such that their excess path lengths are equal. The possible ordered bin values are in the set $\{\delta^{krti}_{l}(0),\ldots,\delta^{krti}_{l}(M-1)\}$ where $M$ is the total number of bins. The RSS measurements for one group of these tuples are seen in the grey box in Fig.~\ref{F:rss_vs_epl} and the histogram of these RSS measurements is shown in Fig.~\ref{F:rss_dist_opt_mix}. We denote the histogram of the RSS measurements whose excess path length is $\delta^{krti}_{l}(m)$ as $\mathbf{h}_{l,m}$ where index $m$ indexes in the set of all excess path lengths.
 We wish to find $b_{l,m}$ such that the mixture model $p(r_l \mid b_{l,m}) = b_{l,m} \cdot p(r_l \mid a) + (1-b_{l,m}) \cdot p(r_l \mid u)$ most closely matches $\mathbf{h}_m$. To do this, we perform 
\begin{equation} \label{E:b_minimize}
b^*_{l,m} = \argmin_{b_{l,m} \in \mathcal{S}_b} \| p(r_l \mid b_{l,m})-\mathbf{h}_{l,m} \|
\end{equation} where $\mathcal{S}_b$ is a set of equally-spaced real valued numbers between $\num{1e-5}$ and 1 and $\| * \|$ is the $\ell^2$-norm. An example optimal mixture model is shown in in Fig.~\ref{F:rss_dist_opt_mix}.

\begin{figure}
\centering
\includegraphics[width=0.38\textwidth]{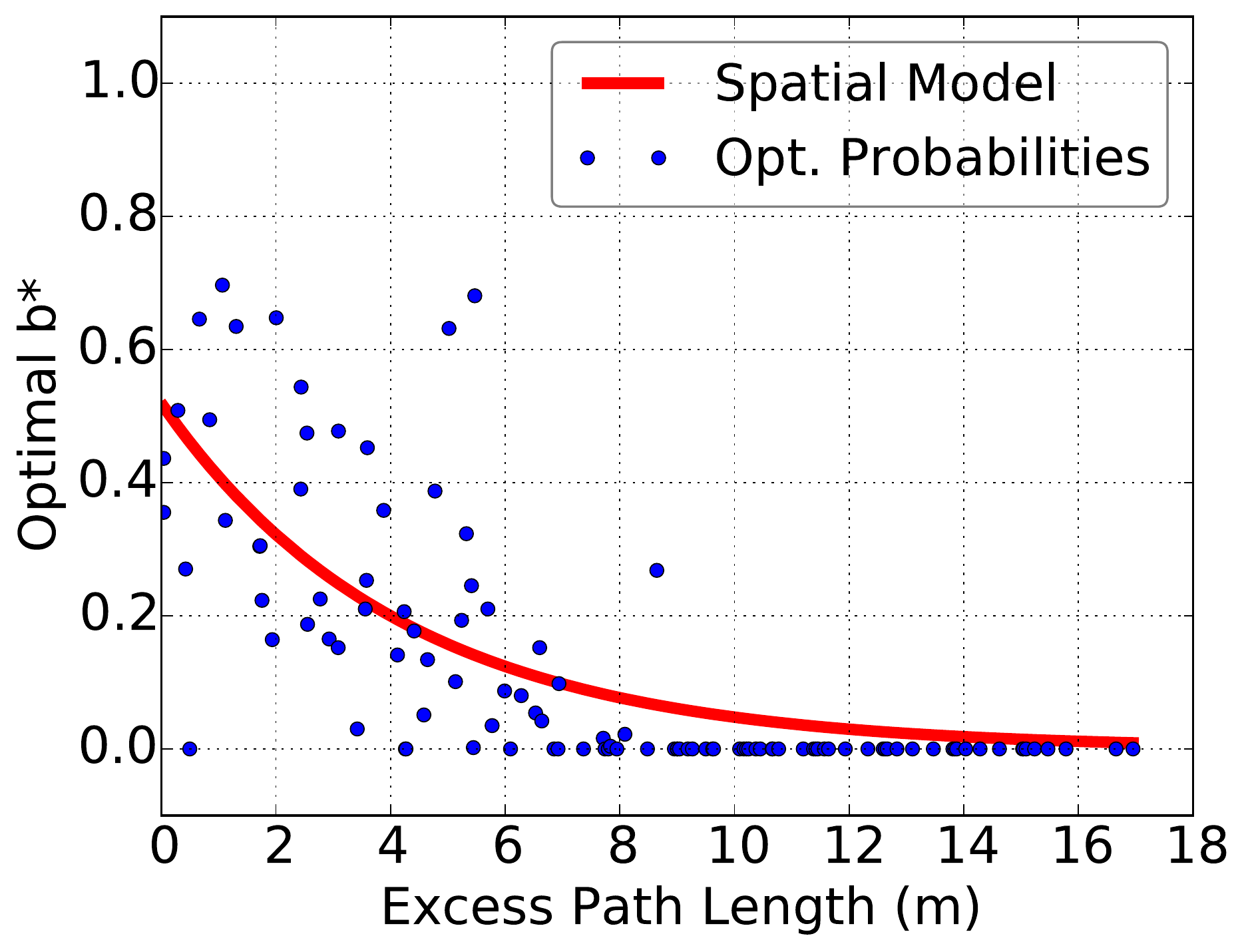}
\caption{Optimal probabilities $b^*$ as a function of excess path length. The estimated spatial model is overlaid.}
\label{F:opt_vs_epl}
\vspace{-1em}
\end{figure}

By performing this process for all excess path length bins, we get the tuples $<\delta^{krti}_{l}(m),b^*_{l,m}>$. We plot these tuples for a link in Fig.~\ref{F:opt_vs_epl}.
 The relationship between $b^*_{l,m}$ and $\delta^{krti}_{l}(m)$ follows our spatial exponential decay function in equation (\ref{E:conditional_probs}). We estimate $\beta_l$ and $\lambda_l$ from a nonlinear least squares solution. The estimation includes constraining $0<\beta_l<1$ to keep the conditional probabilities between 0 and 1 and $\lambda_l>0$ so that a link is always more likely to be affected by a person on or near the link line over a person far from the link line. After $\beta_l$ and $\lambda_l$ are estimated, we do not re-estimate them in this paper and consequently no longer run KRTI. However, re-estimation can be optionally performed at any later time as desired.

%
%
%
%
%
%
%
%
%
%
%
%
%
\subsection{Initial and Transition Probabilities for HMML} \label{S:trans_probs}
A hidden Markov model includes a transition matrix which defines the probabilities of transitioning from one state to another in one time step. In our model, we incorporate the physical constraints of walking inside a building, like walking speed and fixed barriers, into our transition probabilities. To do this, we first label each grid coordinate as either an entrance-exit or as a non-entrance-exit coordinate. Entrance-exits are locations in the area of interest where a person can enter or exit the area of interest. Second, for each grid coordinate, the grid coordinates that are $ \leq 0.75$ m away are labelled as neighbors. For entrance-exit states, we include the out-of-area grid coordinate as a neighbor since the only way to leave the area of interest is via an entrance-exit. For the out-of-area grid coordinate, we label the entrance-exit states as neighbors. However, a grid coordinate cannot be a neighbor if a person must travel through a wall to get to that grid coordinate. Third, we assume that a person is more likely to stay at the current grid coordinate than to transition to another. 

For transition probabilities, the probability of remaining in the same grid coordinate after one time step is set to $0.9$ for all states. For all non-neighbor grid coordinate, we assign a probability $\num{e-200}$. We found that $\num{e-200}$ was the closest value to 0 we could use without encountering numerical representation issues when computing the forward probabilities. This value also gave some, but very little probability, of a target reaching any other state in one time step. For all neighbor grid coordinate, we assign a equal probability so that the sum of probabilities of transitioning from the current grid coordinate to any other grid coordinate equals 1. We note that wall and entrance-exit information is extra information required to create these transition probabilities. Consequently, we will show in Section \ref{S:mpl_eval} how HMML's localization performance is affected if we ignore wall and entrance-exit information.

A hidden Markov model also includes the probability $\pi_k$ that the Markov chain starts in grid coordinate $\mathbf{x}^{grid}_k$. We assume that when the system turns on, the person is located outside of the area of interest with probability $0.95$. All other initial state probabilities are assigned $0.05/P$.

%
%
%
%
%
%
%
%
%
%
%
%
%
\subsection{Continuous Recalibration} \label{S:continuous_recal}
An important element of MPL is that it does not use an empty room calibration period, nor a fingerprint training period, to estimate the mean and variance of the affected and unaffected RSS distributions. Instead, MPL uses unlabelled training data when a person is moving inside a building, a feature we believe adds convenience in deploying a DFL system. Furthermore, MPL is capable of adapting to non-stationary RSS distributions. We enable these features of MPL by running a light weight companion localization method called VRTI \cite{wilson2011vrti}. Using online calibration, VRTI localizes motion by computing the sample variance of a buffer of RSS for each link. The sample variance for each link is used to form an image of the motion, from which we estimate a person's location. We denote the location estimate from VRTI as $\hat{\mathbf{x}}^{vrti}$.

The purpose of running VRTI in tandem with MPL is that VRTI can localize a moving person in spite of a changing environment. With VRTI's location estimate, we not only know where the moving person is located but also where they are \textit{not} located. Here, we are assuming our system is used in a home with a single occupant. If no person is near a link, we can safely update that link's RSS unaffected distribution parameters. We say that $\hat{\mathbf{x}}^{vrti}$ is far from a link if the excess path length of $\hat{\mathbf{x}}^{vrti}$ with respect to link $l$, which we denote $\delta^{vrti}_{l} = d(\mathbf{x}^{vrti},\mathbf{x}^{tx}_l) + d(\mathbf{x}^{vrti},\mathbf{x}^{rx}_l) - d(\mathbf{x}^{tx}_l,\mathbf{x}^{rx}_l)$, is greater than $\delta^{max}_{l}/2$ where $\delta^{max}_{l}$ is the maximum excess path length of any coordinate in $x^{grid}_k$ for $k \in \{0,\ldots,P-1\}$ with respect to link $l$. When $\delta^{vrti}_{l} > \delta^{max}_{l}/2$, we add $r_l$ to the $B$-length FIFO buffer referred to in Section \ref{S:dist_models}.

In as much as VRTI is unable to distinguish between a stationary person and when the area of interest is vacant, the RSS distribution parameters won't be re-estimated when the person is stationary or when a person is outside of the area of interest. However, it is important to update the RSS distributions when the area of interest is vacant. To re-estimate the RSS distribution parameters when the area of interest is vacant, we add $r_l$ to the $B$-length buffer, if it has not been added already, when HMML or MLL says the area of interest is empty, i.e. when $k=P$ is the solution to $\argmax_{0 \leq k \leq P} \alpha_k$ or $\argmax_{0 \leq k \leq P} p_k( \mathbf{r} )$. The mean and variance of the buffer is then used to periodically re-estimate the distribution parameters for both a link's unaffected and affected state as described in Section \ref{S:dist_models}. With both the location estimate of VRTI and HMML or MLL, we are able to perform continuous recalibration without an empty room calibration period. We found that $B=15$ was an appropriate buffer length for our application.

\begin{figure}
\centering
\includegraphics[width=0.38\textwidth]{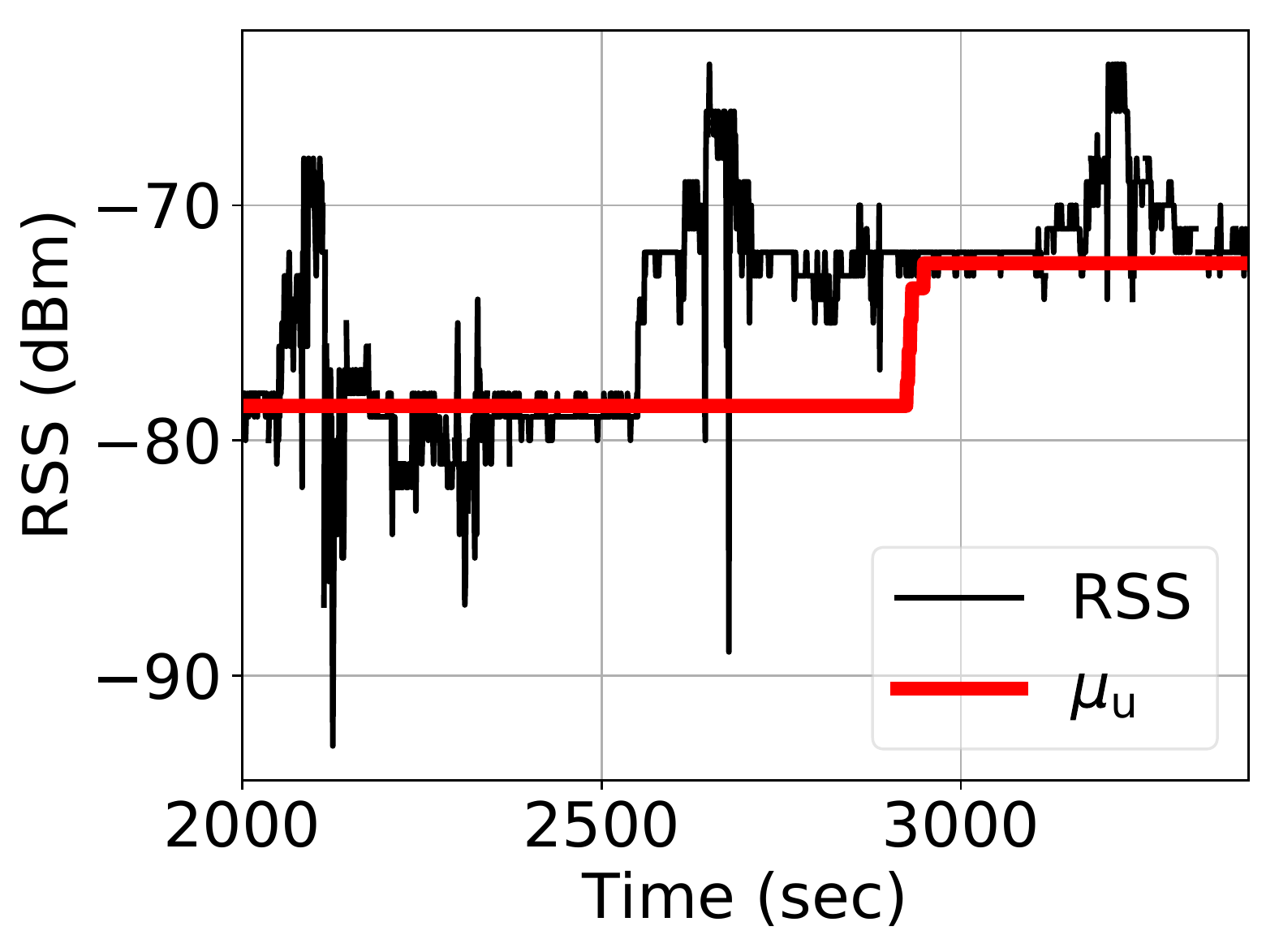}
\caption{Measured RSS on a link before and after a couch is moved at 2550 s. The unaffected RSS mean changes by 6 dBm. The red line shows how our continuous recalibration eventually adjusts the estimate of the unaffected mean after environment changes.}
\label{F:c_recal}
\vspace{-1em}
\end{figure}

As an example of how we perform continuous recalibration, we show in Fig.~\ref{F:c_recal} the measured RSS on a link before and after a couch nearby is moved 15 cm. After the couch is moved at 2550 s, the unaffected RSS increases by 6 dBm. We also show $\mu_u$ for the link during this time period as it is re-estimated. After a few minutes, our unaffected mean RSS estimate adjusts to the increase in RSS due to the changing environment.

%
%
%
%
%
%
%
%
%
%
%
%
%
\subsection{Method Summary}
There are several components to our methods and so we summarize the major points here so that the reader can gain a high level understanding of the algorithm.

The algorithm begins with an unsupervised training period that is performed once. While a person walks around the whole space for several minutes, a history of $\hat{\mathbf{x}}^{krti}$ location estimates and RSS measurements are saved (see Section \ref{S:spatial_param_estimation}.) After the training period is complete, the history of $\hat{\mathbf{x}}^{krti}$ location estimates and RSS measurements are used to estimate $\beta_l$ and $\lambda_l$ for each link (see Section \ref{S:spatial_param_estimation}.) The estimated $\beta_l$ and $\lambda_l$ values are then used to compute $p(s_l = a \mid k)$ and $p(s_l = u \mid k)$ using (\ref{E:conditional_probs}) in Section \ref{S:mixture_model}. Though not part of the unsupervised training, the transition probabilities used in HMML are also computed just once (see Section \ref{S:trans_probs}).

After the unsupervised training has been completed, one of the MPL methods is used to perform location estimation. For every new measurement $\mathbf{r}$, the likelihoods are updated using (\ref{E:mixture_model}) and (\ref{E:likelihoods}) in Section \ref{S:mixture_model}. The likelihoods are then used to update the MLL or HMML location estimate (see Section \ref{S:mixture_model}). In addition to updating the MPL location estimate, a new $\mathbf{r}$ measurement is used to get a new VRTI estimate $\hat{\mathbf{x}}^{vrti}$ (see Section \ref{S:continuous_recal}). Both $\hat{\mathbf{x}}^{vrti}$ and $\hat{\mathbf{x}}^{mpl}$ location estimates are then used to update $\mu_{l,u}, \mu_{l,a}$, $\sigma^2_{l,u}$, and $\sigma^2_{l,a}$ which are used in (\ref{E:log_normal_unaffected}) and (\ref{E:log_normal_affected}) described in Section \ref{S:dist_models}.

%
%
%
%
%
%
%
%
%
%
%
%
%
\subsection{Baseline DFL Methods}
Many DFL methods exist that perform an empty room calibration period or run an online calibration. Empty room calibration is inconvenient for those people who must wait outside of the area of interest. Additionally, DFL methods that require empty room calibration quickly become unreliable estimation methods unless they are frequently recalibrated with empty room snapshots. DFL methods with online calibration lose track of stationary people. Other DFL methods require fingerprint training where a person stands at many locations in the area of interest while RSS measurements are stored in a database. Like DFL methods with empty room calibration, fingerprint DFL becomes unreliable as the real fingerprints diverge from those in the database \cite{mager2015fingerprint}. In this paper, we compare HMML and MLL, both of which address all of these drawbacks, against well-known DFL methods. 

One of these methods is attenuation-based RTI which we refer to as RTI \cite{kaltiokallio2012channels}. RTI requires an empty room calibration where the mean RSS for each link is computed and stored. The absolute difference between $\mathbf{r}$ and the mean RSS is computed and stored as $\mathbf{y}^{rti}$, which is in turn used to compute an image and estimate the person's location. The second method is kernel-based RTI which we refer to as KRTI \cite{zhoa2013krti}. KRTI continuously updates a long and short-term RSS histogram. The kernel distance between these histograms are then computed and stored as $\mathbf{y}^{krti}$, from which the image and the person's location is estimated. For both RTI and KRTI, we use an elliptical model for the weight matrix $\mathbf{W}$ \cite{wilson2010rti}. A regularized-least squares solution is then used to estimate the image $\mathbf{z}$ using the linear relationship $\mathbf{y} = \mathbf{Wz} + \tilde{\mathbf{n}}$ where $\tilde{\mathbf{n}}$ is the noise. The pixels in the image, $\mathbf{z}$, for both RTI and KRTI map to the same grid coordinates $\mathbf{x}^{grid}_k$ for $k=\{0,\ldots,P-1\}$ mentioned in Section \ref{S:mixture_model}. We use the pixel with the greatest value as the location estimate, which we denote $\hat{\mathbf{x}}^{rti}$ for RTI and $\hat{\mathbf{x}}^{krti}$ for KRTI. However, when the image maximum falls below a threshold, we set the location estimate as the out-of-the-area pixel $\mathbf{x}^{grid}_P$. We note that MPL also uses KRTI to estimate spatial parameters, but the baseline KRTI mentioned in this section is separate from KRTI used in MPL. After this point, we distinguish between the two when needed.

The last method is a linear discriminant analysis classifier, which we refer to as LDA, that requires RSS fingerprints at many locations \cite{xu2012classification}. During fingerprinting, a person moves inside of a small area around a known location. The mean RSS of all $L$ links is recorded for fingerprint location index $k^\prime$ and stored as $\boldsymbol{\mu}^{lda}_{k^\prime}$ where $k^\prime=\{0,\ldots,K^\prime\}$, $K^\prime+1$ is the total number of fingerprints. 
The covariance of the RSS over all fingerprint locations, $\hat{\boldsymbol{\Sigma}}$, is then estimated using Ledoit-Wolf shrinkage as $\hat{\boldsymbol{\Sigma}} = (1-\nu)\boldsymbol{\Sigma}^\prime + \nu \rho I$ where $\boldsymbol{\Sigma}^\prime = \sum_{k^\prime=1}^{K^\prime} \sum_{t \in class k^\prime}^{K^\prime}(\mathbf{r}[t]-\boldsymbol{\mu}^{lda}_{k^\prime})(\mathbf{r}[t]-\boldsymbol{\mu}^{lda}_{k^\prime})^T/(T-K^\prime)$ and $T$ is the number of RSS measurement vectors measured during fingerprinting. Ledoit-Wolf shrinkage is a traditional way to estimate a covariance matrix when the number of samples used for estimation is small but the number of variables to estimate is high. We find ourselves in this situation since the number of measurements we record at each fingerprint tends to be small. Finally, we find the $k^\prime$ that maximizes $\mathbf{r}^T \hat{\boldsymbol{\Sigma}}^{-1} \boldsymbol{\mu}^{lda}_{k^\prime} - 0.5 {\boldsymbol{\mu}^{lda}_{k^\prime}}^T \hat{\boldsymbol{\Sigma}}^{-1} \boldsymbol{\mu}^{lda}_{k^\prime}$ which gives us our location estimate $\hat{\mathbf{x}}^{lda}$.

\section{Experimentation}
In this section, we describe the three test sites we used to evaluate the localization performance. We also describe the localization metric used for the evaluation.

\subsection{Test Sites}
\begin{figure*}
  \centering
  \includegraphics[scale=0.38]{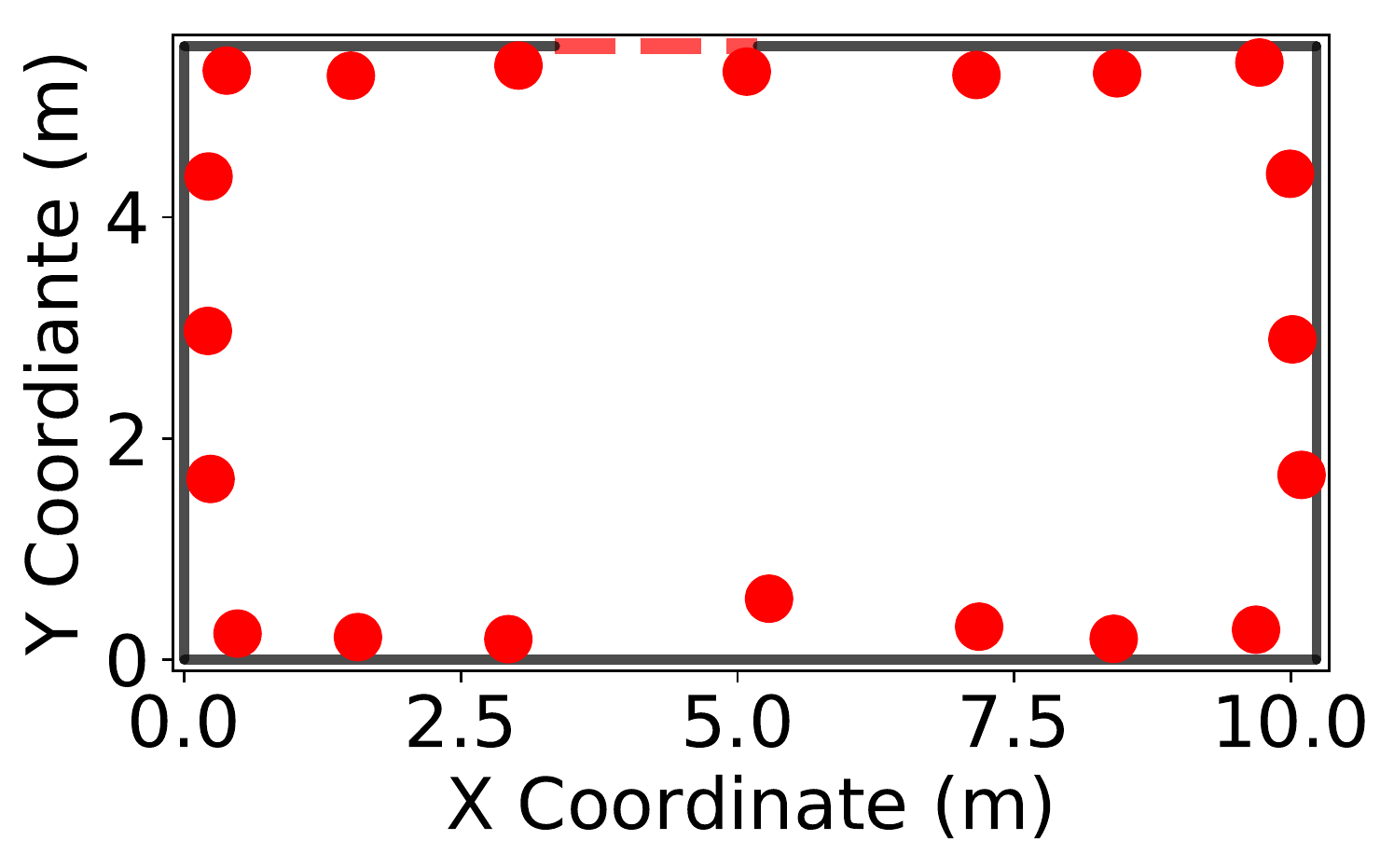}\quad
  \includegraphics[scale=0.38]{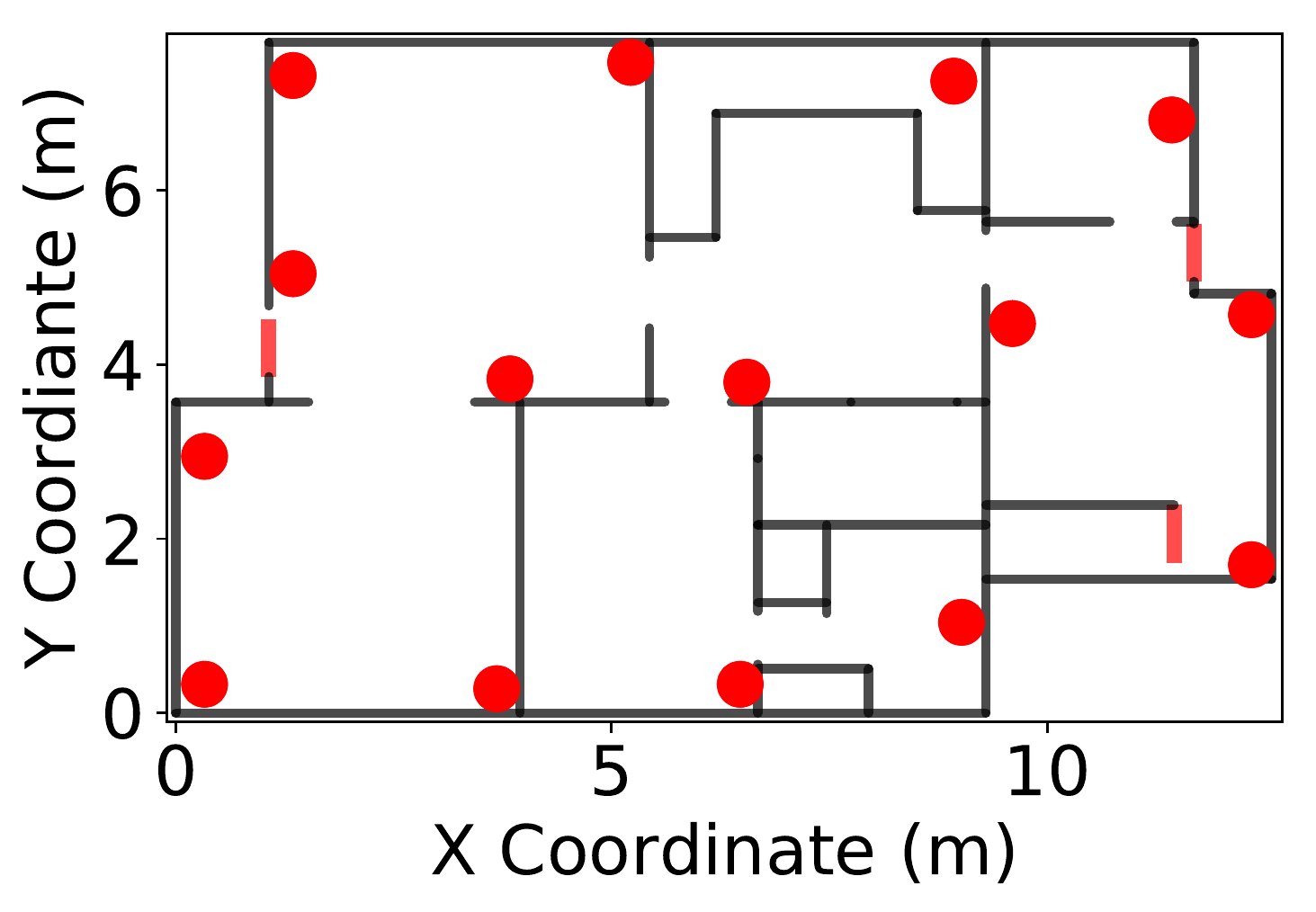}\quad
  \includegraphics[scale=0.38]{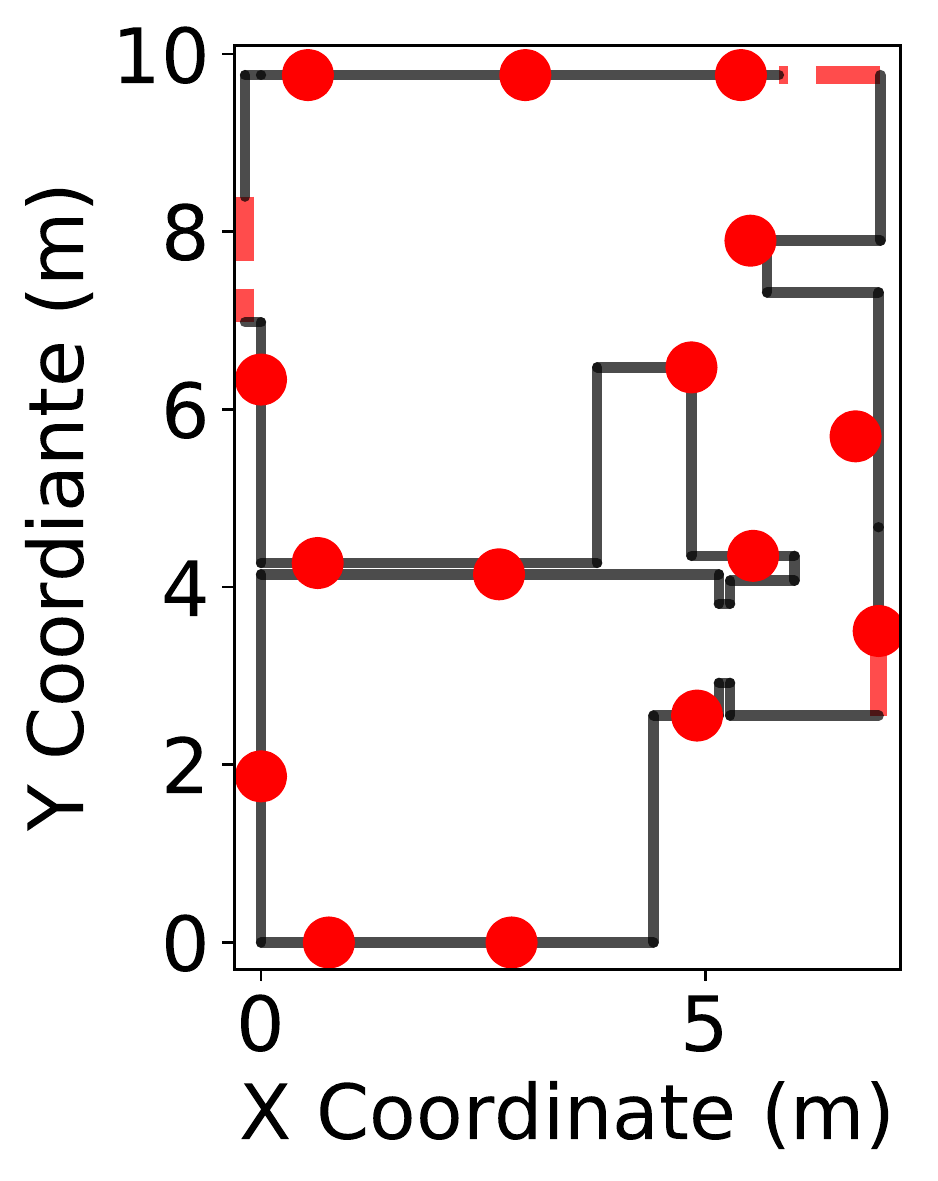}
  \caption{Experiment locations where the walls or barriers are shown in black and the nodes are shown as red circles. The red dashed lines indicate where entrance-exits are located for (left) class room, (middle) first floor, and (right) basement.}
  \label{F:site_locations}
  \vspace{-1em}
\end{figure*}

In our evaluation, we perform experiments at three different test sites. At each site, we first collect a training data set which we use to perform supervised fingerprint training for LDA and to perform unsupervised estimation of $\beta_l$ and $\lambda_l$ for each link in MPL. Additional testing data sets are then performed. Both training and testing data sets include the known location of the person moving through the area. Experimentation at each test site was performed differently, so we describe each test site individually. The floorplans for each site are shown in Fig. \ref{F:site_locations}. For each site, the grid coordinates $\mathbf{x}^{grid}_k$ used for KRTI, RTI, and MPL are generated automatically such that the grid points are evenly distributed around the area of interest. Fingerprint locations for LDA are created only for locations in the area of interest where a person walks during testing.

There are differences in the number of frequency channels measured for each site. We programmed the nodes for two of the sites to use four frequency channels. The data for the third site was collected by other experimenters and they chose to measure eight frequency channels. We chose to measure on four to have a higher sampling rate. In our algorithms, all channels measured are used in the DFL methods.

\subsubsection{Classroom}
Our first test site, which we refer to as site CR, is an empty classroom. We deploy twenty nodes, which measure on four channels, on the inside perimeter of the classroom such that a majority of the links are line of sight. It takes the protocol 0.24 s to get an RSS measurement for each link and channel. 

A total of 30 s of RSS are collected for fingerprints at 100 locations spaced $0.61 m$ apart. At each fingerprint location, the person moved inside a $0.61 m^2$ before moving to the next fingerprint location. The total duration of the training experiment was 55 min. 

During the testing experiments, the room was vacant for the first minute. A person then entered the room and continuously moved to each fingerprint location at least once. A total of twelve test experiments were performed varying between 3 and 12 min in length. 
No objects inside the room were intentionally moved at any time during the training and testing data sets.

\subsubsection{First Floor}
The second test site, which we refer to as site 3F, is the furnished first floor of a home. Thirty nodes are deployed, which measure on eight channels, on the inside perimeter of the house. It takes the protocol 0.82 s to get an RSS measurement for each link and channel. A pair of nodes are attached to a tall stand such that the nodes are 0.3 and 1.3 m above the floor. With the many walls and obstructions in the house, very few of the links are line of sight. 

RSS Fingerprints are collected at thirty-two locations in the house. The total duration of the training experiment was 33 min. During the testing experiments, the house was vacant for the first 50 s. A person then entered the house and moved to the first fingerprint location, standing there for 50 s. After the 50 s elapsed, the person moved to the next fingerprint location where the process continued. The same procedure was followed for the test experiments except that the person stands at each location for 20 s. The duration of each of the 15 test experiments was 10 min.

After the training experiment and each testing experiment, an intentional change to the house was made. For example, a couch was moved, a washer lid was shut, or a sink was filled with water. These intentional changes were performed to simulate the passage of time in a typical house where objects are moved, added, or removed from the area of interest. We note that this training and testing data set was originally created and used in \cite{mager2015fingerprint}.

\subsubsection{Basement Living}
Our last test site, which we refer to as site BL, is a furnished basement. We deploy fifteen nodes, which measure on four channels, on the inside perimeter of the area of interest. It takes the protocol 0.53 s to get an RSS measurement for each link and channel. The floor plan of site BL is shown in Fig.~\ref{F:site_locations}.

During the training experiment, a person continuously moved around the basement at known locations at known times. For fingerprinting, we create several reference locations that serve as the fingerprint location since the person was moving for the duration of the training period. The duration of the training experiment was 15 min. 

During the testing experiments, the basement was vacant for the first minute. A person then entered and continuously walked around the basement. However, the person also reclined on a bed, and sat in an armchair, on a couch, and in a chair for 2 min each at different points during the experiment. This was done to show how DFL methods with online calibration lose track of a stationary person. RSS measurements were recorded for 24 h for seven days. We divide each day into an individual experiment. Each day, a person's ground truth location was recorded for 14 min for performance evaluation. During the seven days, the person performed normal day-to-day tasks and activities, including moving furniture and adding, moving or removing other household items in the area.


\subsection{Localization Accuracy}
The DFL methods we evaluate produce a location estimate for each time $t$. These methods can also indicate that the area of interest is vacant. 
To evaluate each DFL method, we compute the localization error at time $t$ as
\begin{equation}
e[t]  = d(\hat{\mathbf{x}}[t],\mathbf{x}^{true}[t])
\end{equation}
where $\mathbf{x}^{true}$ is the true location coordinate, $\hat{\mathbf{x}}[t]$ is the estimated location coordinate from one of the localization methods, and $d(\hat{\mathbf{x}}[t],\mathbf{x}^{true}[t])$ is the Euclidean distance between the true and estimated location. We then compute the median of $e[t]$ for all $t$ and call it the median Euclidean error $e^{med}$.
\section{Results}
In this section, we discuss the localization performance of MLL and HMML and our baseline DFL methods, RTI, KRTI, and LDA. We show how MLL and HMML outperform all of the baseline methods at three different sites, how MLL and HMML robustly localize moving and stationary people, and how MLL and HMML adapt to changing environments. We then make intentional modifications to MLL and HMML and show how their localization performance is affected.

\subsection{DFL Method Comparison}
\begin{figure}
\centering
\includegraphics[width=0.42\textwidth]{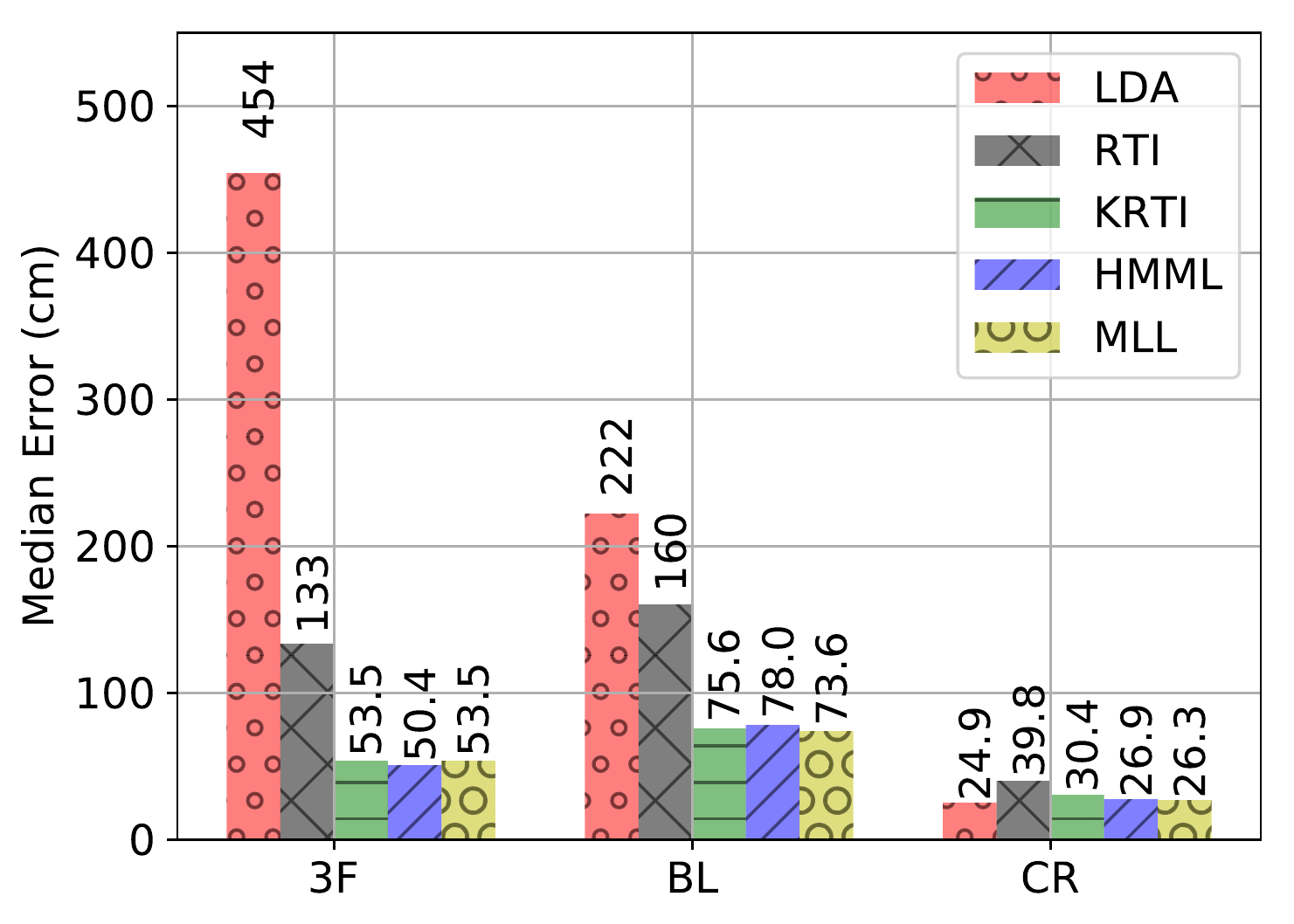}
\caption{Median Euclidean error, $e^{med}$, for LDA, RTI, KRTI, MLL and HMML at three different experiment sites. Only site CR maintained a relatively unchanged environment during the course of the training and testing experiments. 
}
\label{F:method_comare}
\vspace{-1em}
\end{figure}

In this paper, we perform many experiments at three different sites to show how MLL and HMML can accurately localize people in many settings. We show the median error, $e^{med}$, for MLL, HMML, and the baseline DFL methods at the three sites in Fig.~\ref{F:method_comare}.
 For all sites, MLL and HMML outperform or match the baseline methods in localization accuracy. At sites 3F and BL, where there were considerable changes to the environment, MLL and HMML reduce $e^{med}$ by $51\%$ or greater when compared to RTI and LDA. Since the environment changes often at site 3F, RTI and LDA's empty room and fingerprint calibration methods become outdated, resulting in poor performance over time. When the environment does not change, like in the CR experiments, RTI and LDA's localization performance closely matches MLL, HMML and KRTI. However, an unchanging environment, like at site CR, is not likely to exist in most applications. 

In contrast, we observe that MLL reduces the median error by up to $13\%$ and HMML by up to $11\%$ compared to KRTI at the three sites. While this reduction in error seems small, it is important to recognize that it becomes more challenging to make significant reductions in error when the errors are already considerably low given the size of the site areas, the number of nodes deployed, and the spatial diversity of the nodes. Additionally, MLL and HMML have the advantage of localizing stationary people, a feature that is missing in KRTI and other online DFL methods. 

\subsection{Tracking Stationary People Evaluation}
\begin{figure}
\centering
\includegraphics[width=0.42\textwidth]{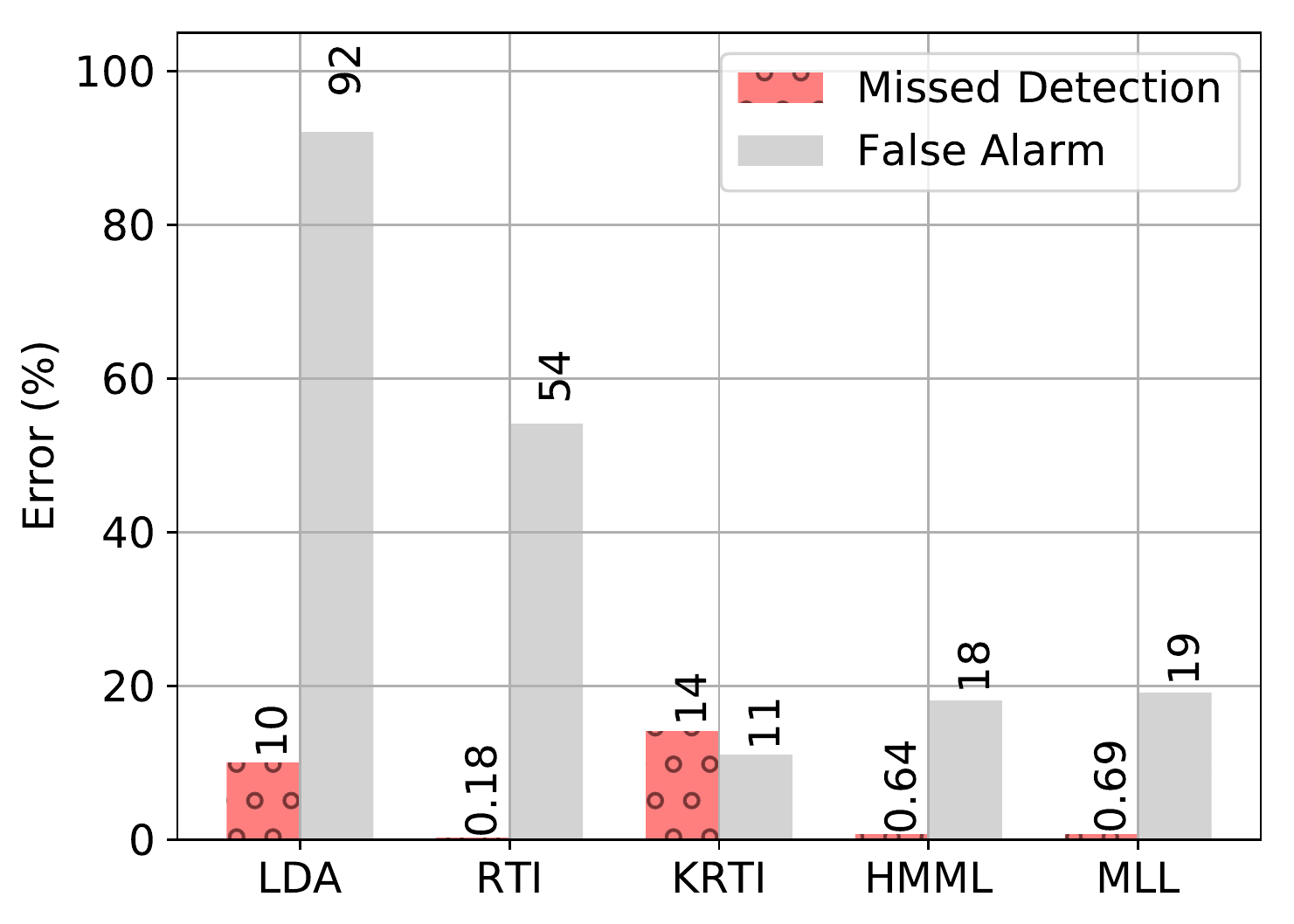}
\caption{The missed detection and false alarm percentages incurred by each DFL method in the BL experiment. Both the HMML and MLL achieve less than $1\%$ missed detection error which is two orders of magnitude lower than what was achieved by KRTI and LDA. HMML and MLL were also able to achieve a false alarm rate that was two or more times as low as LDA and RTI.}
\label{F:sitting}
\vspace{-1em}
\end{figure}

One of the important features of both MLL and HMML is that they keep track of a stationary person. On the other hand, DFL methods that use online calibration only image motion, and so when a person is stationary, these DFL methods eventually lose track of the person. To evaluate each method's ability to localize a stationary target, we show the missed detection and false alarm errors incurred by each method for the BL site in Fig.~\ref{F:sitting}. We show just the results for the BL site since we deliberately included times when the person was stationary during the test.

HMML and MLL both achieve lower missed detection rates than LDA and KRTI by two orders of magnitude. A lower missed detection rate is desirable because it means that the method is better able to localize a stationary target. Interestingly, RTI achieved the lowest missed detection rate. However, it also suffered from a $54\%$ false alarm rate which is more than two times as great as MLL and HMML's rate. Overall, MLL and HMML achieve the best missed detection rate without sacrificing on their false alarm rate. 

We note that the false alarm rates for all five DFL methods are very high. One reason for this is because there were few samples when the person was not in the area of interest. As such, falsely detecting even a few of those samples as presence raised the false alarm rate. Had we collected more samples during which the space was vacant, we anticipate that the false alarm percentages would reduce significantly. A second reason why the false alarm rates are high in general for each DFL method is becuase we chose to weight the risk of loosing track of a stationary person higher than falsely detecting a person's presence. We also desired to have a low localization error. Both of these factors influenced parameter selection and naturally increased the false alarm rate. However, we felt these design choices would be useful in elder care applications when it is more important to keep track of where a person is inside a space while allowing for some false alarms.

\subsection{Continuous Recalibration Evaluation}
\begin{figure}
\centering
\includegraphics[width=0.42\textwidth]{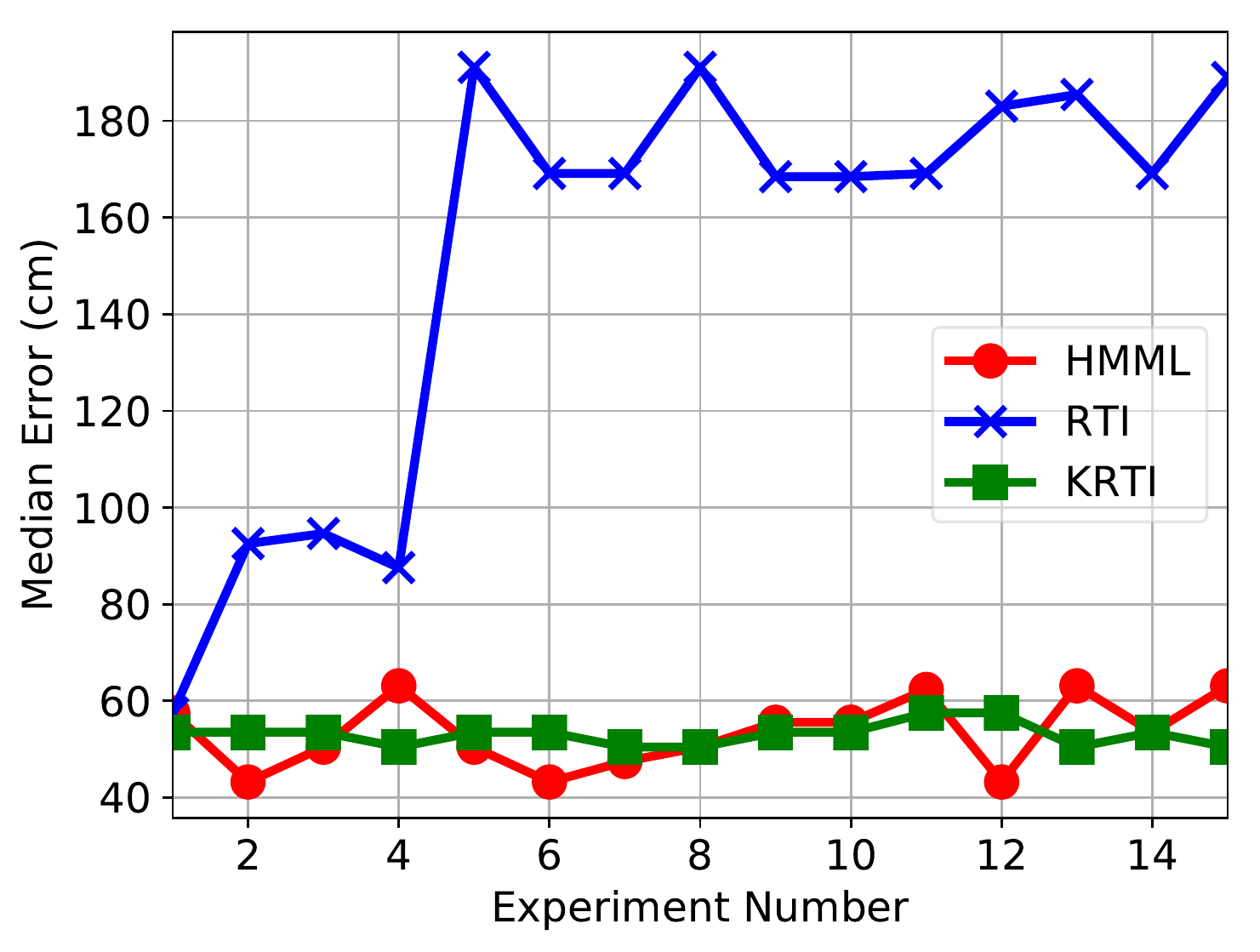}
\caption{Median error for each of the sixteen test experiments performed at 3F for RTI, KRTI and HMML. Intentional changes to the environment were made after each experiment. HMML adjusts to these changes with continuous recalibration. KRTI naturally adjusts to the changes since this method localizes motion. RTI, which performs offline calibration, gradually suffers in localization performance as the empty room calibration measurements diverge from measurements made when the system turned on.}
\label{F:environ_change_brad}
\vspace{-1em}
\end{figure}

\begin{figure}
\centering
\includegraphics[width=0.42\textwidth]{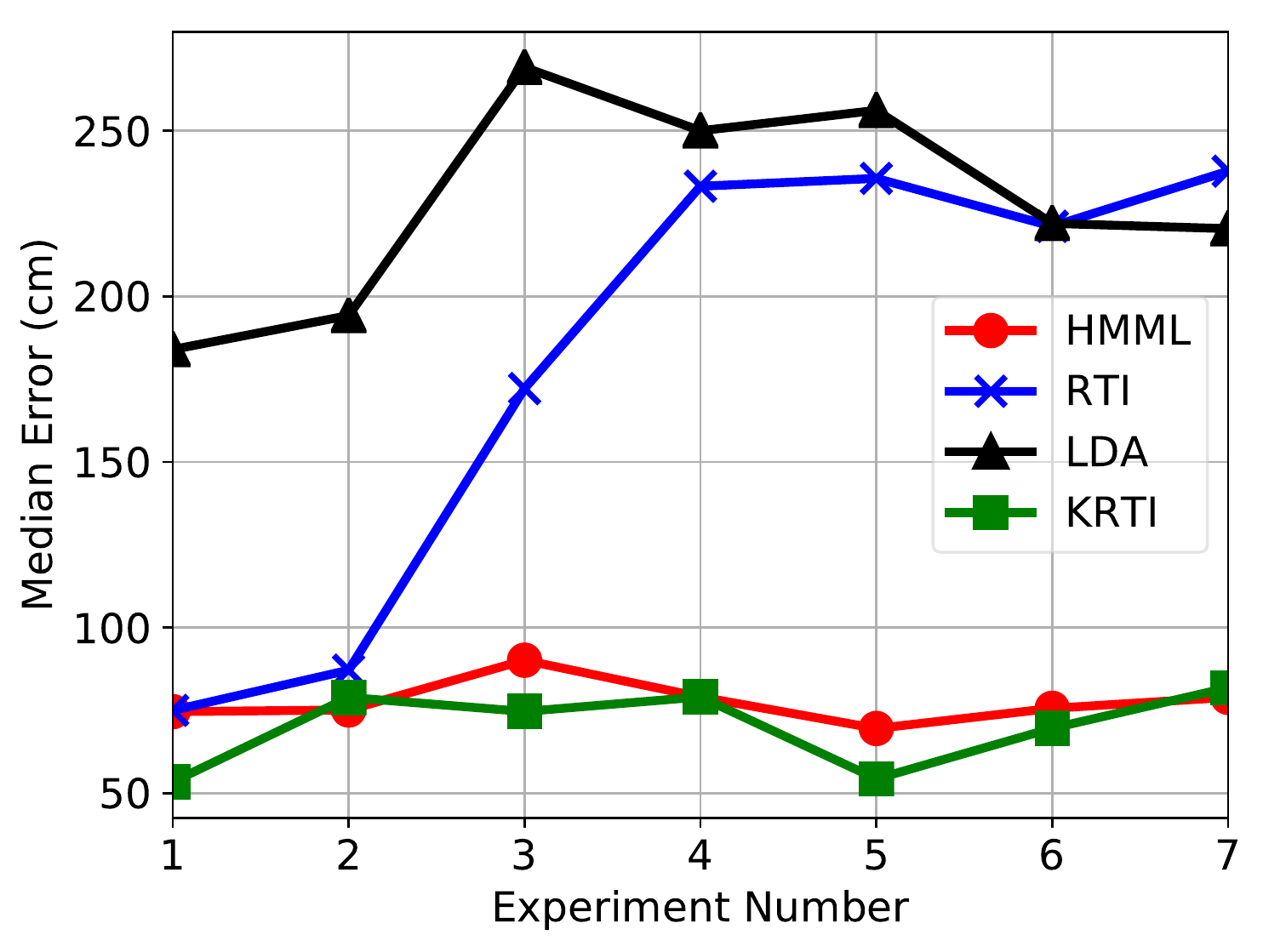}
\caption{Median error for each of the seven test experiments performed at BL for LDA, KRTI, RTI, and HMML. Changes to the environment after each test were the result of day-to-day living. HMML adjusts to these changes with continuous recalibration. RTI, which performs empty room calibration, and LDA, which performs a fingerprint calibration, gradually suffers in localization performance as the empty room and fingerprint calibration measurements become out dated.}
\label{F:environ_change_jean}
\vspace{-1em}
\end{figure}

Another important feature of MLL and HMML is that it can robustly localize a person even in changing environments. DFL methods that perform an empty room or fingerprint training will gradually suffer in localization performance. As the environment changes, the empty room calibration measurements gradually diverge from those measurements recorded during training. We show an example of localization performance in changing environments in Fig.~\ref{F:environ_change_brad}. The median error is shown for RTI, KRTI and HMML for each of the fifteen test experiments at site 3F. Again, MLL's performance is similar to HMML, and we therefore did not want to clutter the figure by including MLL's results. We exclude LDA's results since its large errors make it difficult to see the differences between HMML and KRTI.
 HMML, KRTI, and RTI perform equally as well for the first experiment since RTI's empty room calibration measurements are current. With each successive experiment, intentional changes are made to the environment. As a result, RTI's localization error increases, even doubling by experiment five. Without frequent empty room calibration, RTI is unable to provide a reliable location estimate. On the other hand, HMML, as well as KRTI, robustly localizes the person in spite of a changing environment. With the addition of LDA, these same observations can be seen in Fig.~\ref{F:environ_change_jean} which shows the median error for each of the seven test experiments at site BL.

\subsection{MPL Feature Evaluation} \label{S:mpl_eval}
\begin{figure}
\centering
\includegraphics[width=0.42\textwidth]{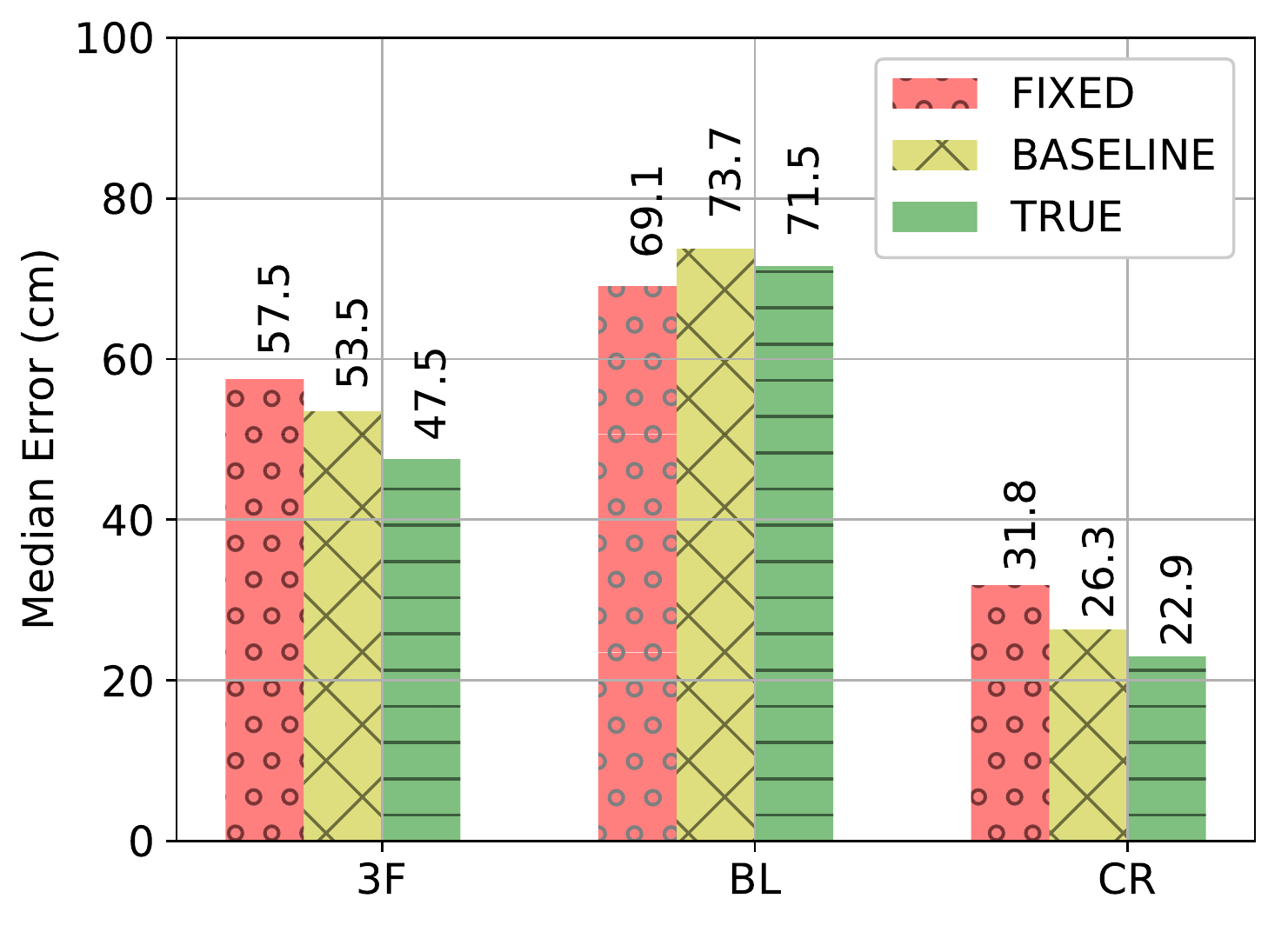}
\caption{Median error for MLL at three different experiment sites. In FIXED, we use the same $\lambda$ and $\beta$ parameters for all links. In TRUE, the true location $\mathbf{x}^{true}$ is used instead of $\mathbf{x}^{krti}$ to estimate $\lambda$ and $\beta$. BASELINE indicates no modification.}
\label{F:mle_compare}
\vspace{-1em}
\end{figure}

\begin{figure}
\centering
\includegraphics[width=0.42\textwidth]{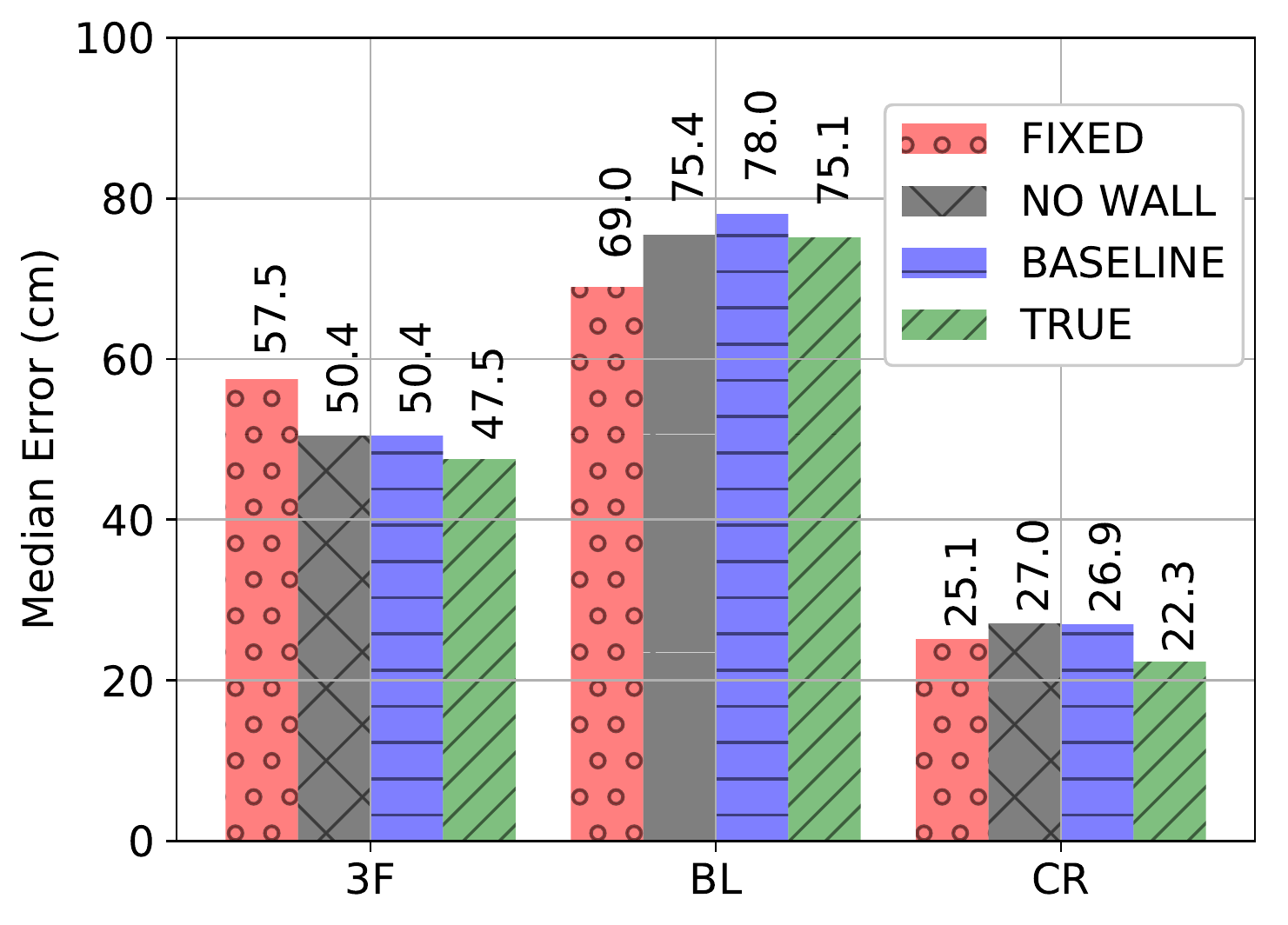}
\caption{Median error for a modified HMML at three different experiment sites. In FIXED, we use the same $\lambda$ and $\beta$ parameters for all links. In NO WALL, we ignore wall and entrance-exit information when creating the transition probabilities. BASELINE is HMML without modifications. In TRUE, the true location $\mathbf{x}^{true}$ is used instead of $\mathbf{x}^{krti}$ to estimate $\lambda$ and $\beta$.}
\label{F:hmm_compare}
\vspace{-1em}
\end{figure}

In this section, we intentionally modify parts of MLL and HMML to see how localization is affected. We make the following four modifications to HMML and MLL.
\begin{itemize}
    \item First, we fix $\lambda$ and $\beta$ for all links instead of estimating them. The values for $\lambda$ and $\beta$ are set to achieve the lowest localization error. We call this modification FIXED. We perform this modification for both HMML and MLL.
    \item Second, we use the true location $\mathbf{x}^{true}$ instead of $\mathbf{x}^{krti}$ to estimate $\lambda$ and $\beta$ in the spatial model parameter estimation block in Fig.~\ref{F:block_diagram}. We call this modification TRUE and make the modification for both MLL and HMML.
    \item Third, we ignore wall and entrance-exit information when creating the transition probabilities. We call this modification NO WALL but only apply this to HMML since MLL does not use transition probabilities.
    
\end{itemize} The unmodified MLL and HMML we call BASELINE. Only one modification is made to MLL and HMML at a time. For each modification, we perform localization using the data from each site and show the results for MLL in Fig.~\ref{F:mle_compare} and for HMML in Fig.~\ref{F:hmm_compare}.

With the FIXED modification, we entirely eliminate the KRTI and parameter estimation blocks, as seen in Fig.\ref{F:block_diagram}, from MPL. In their place, $\lambda$ and $\beta$ are tuned by the user and, consequently, MLL and HMML are ready to run when the system starts. There is no need for any calibration. As seen in Fig.~\ref{F:hmm_compare}, setting all spatial parameters to be the same value for all links increases the median error for HMML by 7 cm at site 3F, reduces the error by 1 cm at site CR, and reduces the error by 9 cm at site BL when compared to BASELINE. From Fig.~\ref{F:mle_compare}, we observe that setting all spatial parameters to be the same value for all links increases the median error for MLE by 4 cm at site 3F, 5 cm at site CR, and reduces the median error by 4 cm at site BL when compared to BASELINE. 

Our first observation is that there are cases when performing a calibration for MLL and HMML actually lead to poorer performance than if a fixed set of parameters were applied. We suspect that there are environments where the estimated location provided by the KRTI block of MPL has a high error which leads to less accurate estimated system parameters. The convenience of the unsupervised calibration may be worth a slight loss in localization accuracy.

With the TRUE modification, we require the user to provide labelled RSS data with their true location during the training phase. We wish to see if having labelled training data would improve the estimation of the spatial parameters, and, in turn, improve localization. From Fig.~\ref{F:hmm_compare} and Fig.~\ref{F:mle_compare}, we observe that at all three sites, supplying labelled RSS data matches or decreases localization error for MLL and HMML when compared to BASELINE. The improvement to localization performance can be as great as 6 cm. The user may decide that the performance gains are too small to warrant having to provide labelled RSS data. However, the trade off is the additional computational overhead needed to run KRTI for the location estimates.

Finally, with the NO WALL modification, we eliminate the need for extra wall information to be entered into HMML prior to operation. From Fig.~\ref{F:hmm_compare}, we observe that the localization error increases at most by 0.1 cm and decreases by 2.5 cm when we ignore wall information. This suggests that there is really no significant loss or gain by including wall information into the transition probabilities. As a result, we can save in some overhead cost in computing the transition probabilities without risking loss in localization accuracy.

\subsection{Complexity and Feature Trade Offs}
We have shown how MML and HMML achieve a lower localization error than other DFL methods and can do so in a changing environment and without an empty room or fingerprint calibration period. Other DFL methods do not share all of these same properties. The trade off, however, for using MLL and HMML over other DFL methods is the greater memory and computational complexity required to run them. In Table~\ref{T:feature_complexity}, we compare the properties of MLL, HMML and other DFL methods, their calibration requirements, and their memory and computational complexity.
\begin{table*}
\centering
\caption{Features offered by DFL methods along with their calibration requirements and memory and computational complexity. For reference, $L$ is the number of RSS measurements, $P$ is the number of grid coordinates, and $R$ is the number of bins used for histograms. The costs for continuously running VRTI for HMML and MLL are included in their complexity.}
\label{T:feature_complexity}
\begin{tabular}{cccccccccc}
 & \multicolumn{4}{c}{Initial Calibration} &  & \multicolumn{2}{c}{Localize} & \multicolumn{2}{c}{Complexity} \\ \cline{2-5} \cline{7-10} 
\multicolumn{1}{c|}{\begin{tabular}[c]{@{}c@{}}DFL\\ Method\end{tabular}} & \begin{tabular}[c]{@{}c@{}}Labelled\\ Empty\end{tabular} & \begin{tabular}[c]{@{}c@{}}Labelled\\ Fingerprint\end{tabular} & \begin{tabular}[c]{@{}c@{}}Unlabelled\\ Occupied\end{tabular} & \multicolumn{1}{c|}{None} & \multicolumn{1}{c|}{\begin{tabular}[c]{@{}c@{}}Constant perf.\\ in changing env.'s\end{tabular}} & Motion & \multicolumn{1}{c|}{\begin{tabular}[c]{@{}c@{}}Stationary\\ Person\end{tabular}} & Memory & Computation \\ \hline
\multicolumn{1}{c|}{LDA} &  & X &  & \multicolumn{1}{c|}{} & \multicolumn{1}{c|}{} & X & \multicolumn{1}{c|}{X} & $LP + P$ & $LP + P$ \\ \hline
\multicolumn{1}{c|}{KRTI} &  &  &  & \multicolumn{1}{c|}{X} & \multicolumn{1}{c|}{X} & X & \multicolumn{1}{c|}{} & $LP + 2LR$ & $LR + LP$ \\ \hline
\multicolumn{1}{c|}{RTI} & X &  &  & \multicolumn{1}{c|}{} & \multicolumn{1}{c|}{} & X & \multicolumn{1}{c|}{X} & $LP$ & $L + LP$ \\ \hline
\multicolumn{1}{c|}{HMML} &  &  & X & \multicolumn{1}{c|}{} & \multicolumn{1}{c|}{X} & X & \multicolumn{1}{c|}{X} & $2LP + 2LR + P + P^2$ & $L+2LP + P^2$ \\ \hline
\multicolumn{1}{c|}{MLL} &  &  & X & \multicolumn{1}{c|}{} & \multicolumn{1}{c|}{X} & X & \multicolumn{1}{c|}{X} & $2LP + 2LR$ & $L+2LP$ \\ \hline
\end{tabular}
\vspace{-1em}
\end{table*} 

The table shows how each of the DFL methods we compare in this paper use different initial calibration methods. RTI requires a period of time when the area of interest is vacant while LDA requires fingerprint training. Empty room calibration and fingerprint training may be feasible, but for aging in place applications, these calibration methods would cause too much inconvenience by having to frequently recalibrate or retrain them. MPL, on the other hand, is calibrated by a person moving around the area of interest, which is likely something that the person would be doing anyway during the course of a day. The big difference between MPL and RTI and LDA is that MPL is able to achieve constant localization performance in changing environments whereas RTI and LDA cannot do so unless empty room calibration or labelled fingerprint training measurements are frequently performed.

In contrast, KRTI does not require any calibration and achieves constant performance in changing environments. The trade off is that KRTI and other online calibration DFL methods are unable to localize stationary people. Applications like home automation and assisted living are dependent on knowing where a person is, even when they are stationary. MPL is able to localize a stationary person, but the trade off is that MPL must be calibrated by having a person walk around in the area of interest. Unlike \cite{xu2012classification}, this movement does not need to be labelled with location data. Additionally, we showed that by modifying MPL to use tuned values for $\lambda$ and $\beta$ instead of estimating them, we would also classify MPL as requiring no calibration like KRTI.

A significant trade off to consider when using MLL or HMML are their relatively higher memory and computational complexity compared to other DFL methods. We observe that, compared to other DFL methods, HMML has an extra $P^2$ memory factor which is used to store the transition probabilities and an extra $P^2$ term in computational complexity which is needed to compute the forward algorithm. The greater memory and computational cost of HMML were used to add a temporal component to localization. We set out to see if the temporal properties of HMML would provide greater localization accuracy than MLL, but we did not observe those gains. HMML is therefore at a disadvantage when $P$ is increased when compared to any of the DFL methods we compared.

The alternative to HMML is MLL, which does not include the extra $P^2$ memory and $P^2$ computation like for HMML. We also saw in the previous sections that MLL often performed localization just as well as HMML. Since MLL ignores the temporal component that HMML embraces, it reduces the computational and memory cost by a nontrivial amount. One question to be asked though is, why doesn't HMML benefit from the addition of transition probabilities? We note that both MLL and HMML use the likelihood probabilities in equation (\ref{E:likelihoods}) which turn out to be values either very near 0 or very near 1 since $L$ is large for all of our experiments. Therefore, the transition probabilities play an insignificant role when inductively computing the joint probabilities $\alpha_k$. The localization results of MLL and HMML demonstrate that, at least in the experiments we performed, there is no clear advantage for including temporal properties into the localization problem by using HMML. In general, estimating with temporal properties is helpful since it smooths unlikely jumps in the location estimate. However, in the experiments we performed, the measurement dimension was so high that it resulted in likelihoods that overwhelmed any contribution the transition probabilities could make in HMML. Had this not been the case, and had we been able to sample more frequently, HMML may have been able to reduce the number of big jumps in the location estimate.

If we were to consider using another DFL method other than MLL, KRTI would be a smart choice. KRTI has no need of calibration and it is highly accurate. However, MLL, which only adds another $LP$ of memory and $LP$ of computation by comparison, buys the advantage to track stationary people while achieving similar localization performance. This advantage is an important feature to have when monitoring an aging, home-bound family member or patient and for enabling smart home features that depend on sensing presence.



\section{Conclusion}
In this paper, we have presented a new signal strength-based Bayesian device-free localization system called model-based localization (MPL) that can localize stationary people, does not require an empty room calibration period, and achieves constant localization performance in changing environments. We developed a new mixture model where the probability of a person occupying a location is a function of signal strength measurements from a wireless sensor network. 
Our mixture model allowed for uncertainty in the state of the link as a function of the person's location. We developed two realizations of MPL including MLL and HMML which compute the probabilities of a person's location based on the RSS measurements observed. MLL computes the likelihoods of observing RSS measurements given a person's location while HMML computes the joint probabilities of observing RSS measurements given a person's location using the forward algorithm of a hidden Markov model. We also developed a method to continuously recalibrate our model to a changing environment. 

To validate the performance of MLL and HMML, we performed a series of experiments at three different sites and compute the localization error of MLL, HMML and three other DFL methods. We demonstrated that MLL and HMML outperform the baseline methods in terms of localization accuracy, that MLL and HMML are capable of localizing a stationary person when other baseline methods cannot, and that MLL and HMML achieves constant localization performance even when the environment changes. In addition, we demonstrated that the MLL can perform localization as well HMML but with a fraction of the memory and computational cost. For assisted living and home automation applications, MPL offers an important advantage of constant localization performance and tracking stationary people without significant costs in computational complexity, memory usage, or convenience.
\ifCLASSOPTIONcompsoc
  \section*{Acknowledgments}
\else
  \section*{Acknowledgment}
\fi
This material is based upon work supported in part by the National Science Foundation under grant \#1407949 and the Army Research Office under grant \#69215CS.  

\bibliographystyle{IEEEtran}
\bibliography{ref_list}

\begin{IEEEbiography}[{\includegraphics[width=1in,height=1.25in,clip,keepaspectratio]{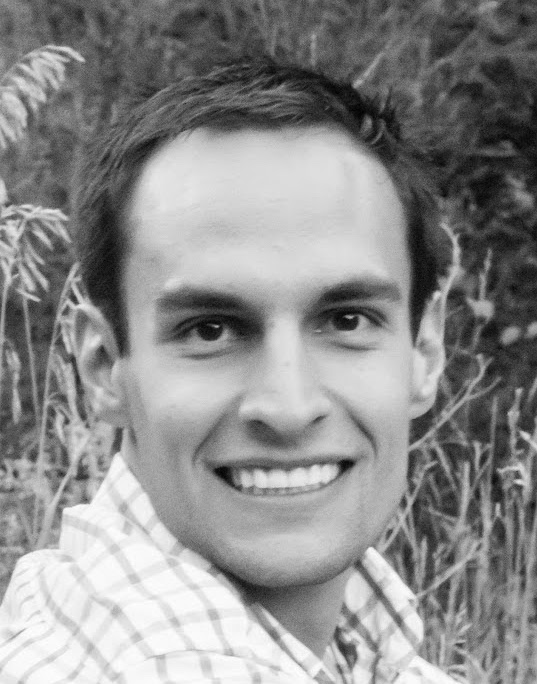}}]{Peter Hillyard}
Peter Hillyard received the B.S., M.E., and Ph.D.
degrees from the Department of Electrical and Computer Engineering, University of Utah. He completed a postdoctoral research position in the Sensing and Processing Across Networks (SPAN) Lab at the University of Utah in wireless RF-based respiratory monitoring technology and now works at Xandem Technology as a Senior Engineer. His  research  interests  include  signal processing, estimation and detection theory, and pattern classification.
\end{IEEEbiography}

\begin{IEEEbiography}[{\includegraphics[width=1in,height=1.25in,clip,keepaspectratio]{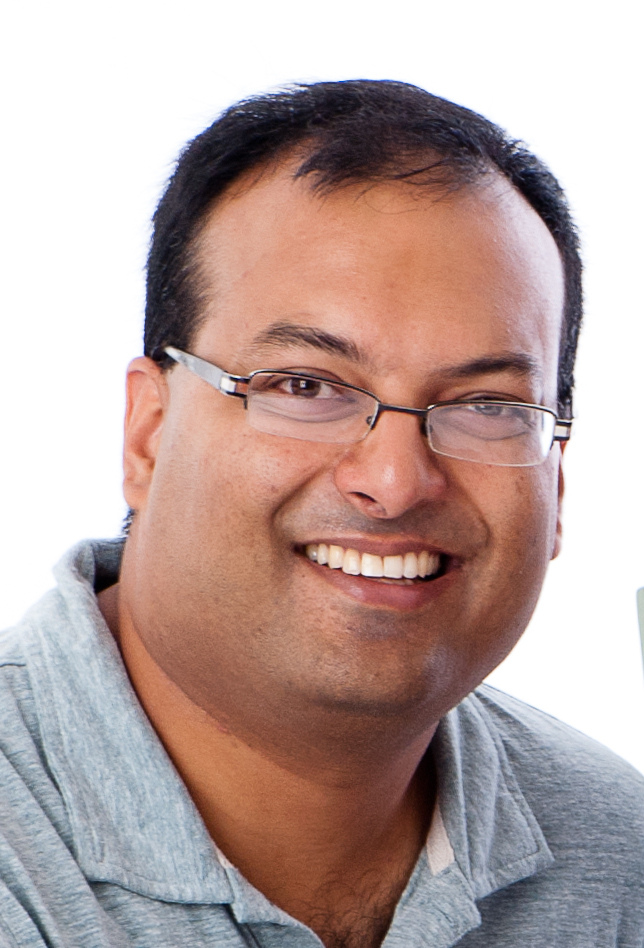}}]{Neal Patwari}
is a Professor in the Department of Electrical and Computer Engineering at the University of Utah, with an adjunct appointment in the School of Computing.  He directs the Sensing and Processing Across Networks (SPAN) Lab, which performs research at the intersection of statistical signal processing and wireless networking, for improving wireless sensor networking and for RF sensing, in which the radio interface is the sensor.  His research perspective was shaped by his BS and MS in EE at Virginia Tech, his research work at Motorola Labs in Plantation, Florida, and his Ph.D. in EE at the University of Michigan. He received the NSF CAREER Award in 2008, the 2009 IEEE Signal Processing Society Best Magazine Paper Award, and the 2011 U.\ of Utah Early Career Teaching Award.  He has co-authored papers with best paper awards at IEEE SenseApp 2012 and at the ACM/IEEE IPSN 2014 conference.  Neal has served on technical program committees for IPSN, MobiCom, SECON, and SenSys.
\end{IEEEbiography}

\end{document}